\documentclass[12pt, a4paper,doublespace]{article}

\usepackage{caption}
\usepackage{subcaption}
\usepackage[noae]{Sweave}
\usepackage{url}
\usepackage{color}
\usepackage{amsthm,amssymb,amsmath}
\usepackage[square,sort,comma,numbers]{natbib}

\let\code=\texttt
\let\proglang=\textsf
\newcommand{\pkg}[1]{{\fontseries{b}\selectfont #1}}

\begin{document}

\title{$K$-Adaptive Partitioning for Survival Data, with an Application to Cancer Staging}

\author{Soo-Heang Eo\footnote{Department of Statistics, Korea University, Seoul, Korea}       \and
	Hyo Jeong Kang\footnote{Department of Pathology, Asan Medical Center, Seoul, Korea}  \and
        Seung-Mo Hong\footnote{Corresponding author, Department of Pathology, University of Ulsan College of Medicine,  Asan Medical Center, 88 Olympic-ro 43-gil, Songpa-gu, Seoul, Korea, 138-736, smhong28@gmail.com} \and
	HyungJun Cho\footnote{Corresponding author, Department of Statistics, Korea University, Anam-dong, Seongbuk-gu, Seoul, Korea, 136-701,  hj4cho@korea.ac.kr} 
}

\maketitle

\begin{abstract}
In medical research, it is often needed to obtain subgroups with heterogeneous survivals, which have been predicted from a prognostic factor.
For this purpose, a binary split has often been used once or recursively; however, binary partitioning may not provide an optimal set of well separated subgroups.
We propose a multi-way partitioning algorithm, which divides the data into $K$ heterogeneous subgroups based on the information from a prognostic factor. 
The resulting subgroups show significant differences in survival.
Such a multi-way partition is found by maximizing the minimum of the subgroup pairwise test statistics.
An optimal number of subgroups is determined by a permutation test. 
Our developed algorithm is compared with two binary recursive partitioning algorithms. 
In addition, its usefulness is demonstrated with a real data of colorectal cancer cases from the Surveillance Epidemiology and End Results program.
We have implemented our algorithm into an \proglang{R} package \pkg{kaps}, which is freely available in the Comprehensive R Archive Network (CRAN).


\end{abstract}

\section{Introduction}
Clinicians are interested in obtaining a handful of subgroups or stages with heterogeneous survivals by partitioning a prognostic factor for prognostic diagnosis \cite{schumacher2006prognostic}.  
The tumor node metastasis (TNM) staging system is the most widely used cancer staging system, and provides critical information about prognosis and about estimation for responsiveness to specific treatment for cancer patients \cite{SEER}. 
The TNM staging system is composed of three classifications: T classification based on the extent or size of the primary tumor, N classification determined by the involvement of the regional lymph nodes (LNs), and M classification by distant metastasis. 
Each T, N, or M classification is decided by grouping cases with similar prognosis. 
When T classification, based solely on the size of the primary tumor such as breast cancer, or N classification, in several gastrointestinal tract cancers, was determined, an increased tumor size or increased number of metastatic regional LNs is linked with a worse prognosis of cancer patients.

\begin{figure}
\centering
\includegraphics[width=14cm,trim = .1 50 .1 70, clip]{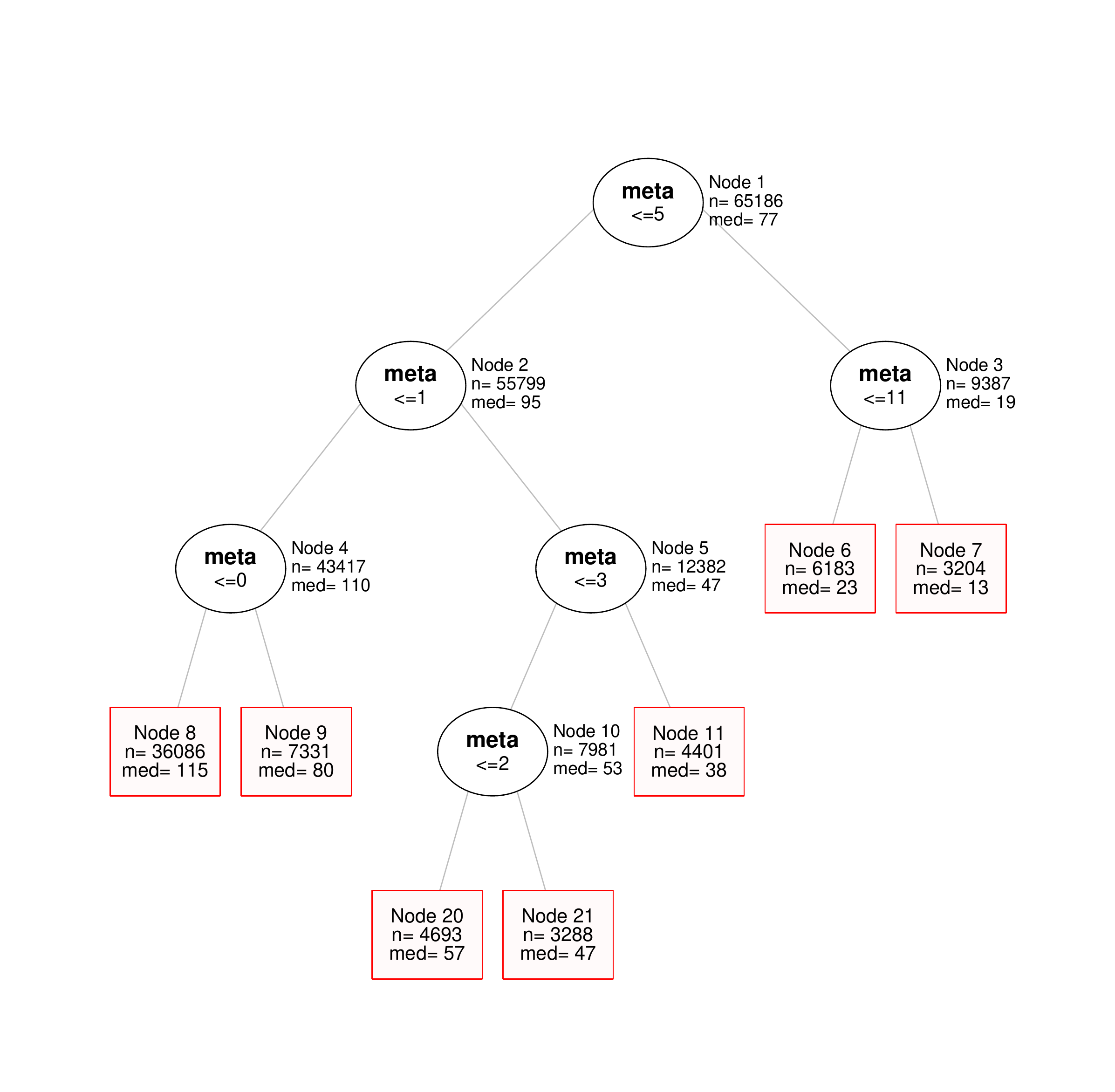}
\caption{Tree diagram from the log-rank survival tree for the colorectal cancer data. Each oval including a split rule depicts an intermediate node and each rectangle with the node number (Node), the number of observations (n), and the median survival time (med) describes a terminal node. An observation goes to the left subnode if and only if the condition is satisfied. Information related to the intermediate node is presented on the right side of the ovals. }
\label{colon:lrtree}
\end{figure}

Such a staging system can be constructed by various partitioning techniques. 
Several studies have been conducted for partitioning a prognostic factor \cite{hilsenbeck1996practical, contal1999application, hothorn2003, Williams06finding, ishwaran2009novel, baneshi2011dichotomisation}.  
Various test statistics have been employed to obtain subgroups with different survivals or cancer stages; however, these approaches only revealed two subgroups such as low- and high-risk patients \cite{altman1994dangers, heinzl2001cautionary}. 
Tree-structured or recursive partitioning methods were also utilized to find an optimal set of cutpoints \cite{lausen1994classification, hong2007}, so as to obtain several heterogeneous subgroups. 
Binary recursive partitioning selects the best point at the first split, but its subsequent split points may not be optimal in combination. 
Some subgroups differ substantially in survival, but others may differ barely or insignificantly. 

\begin{figure}
  \begin{center}
  \includegraphics[width= .7\textwidth]{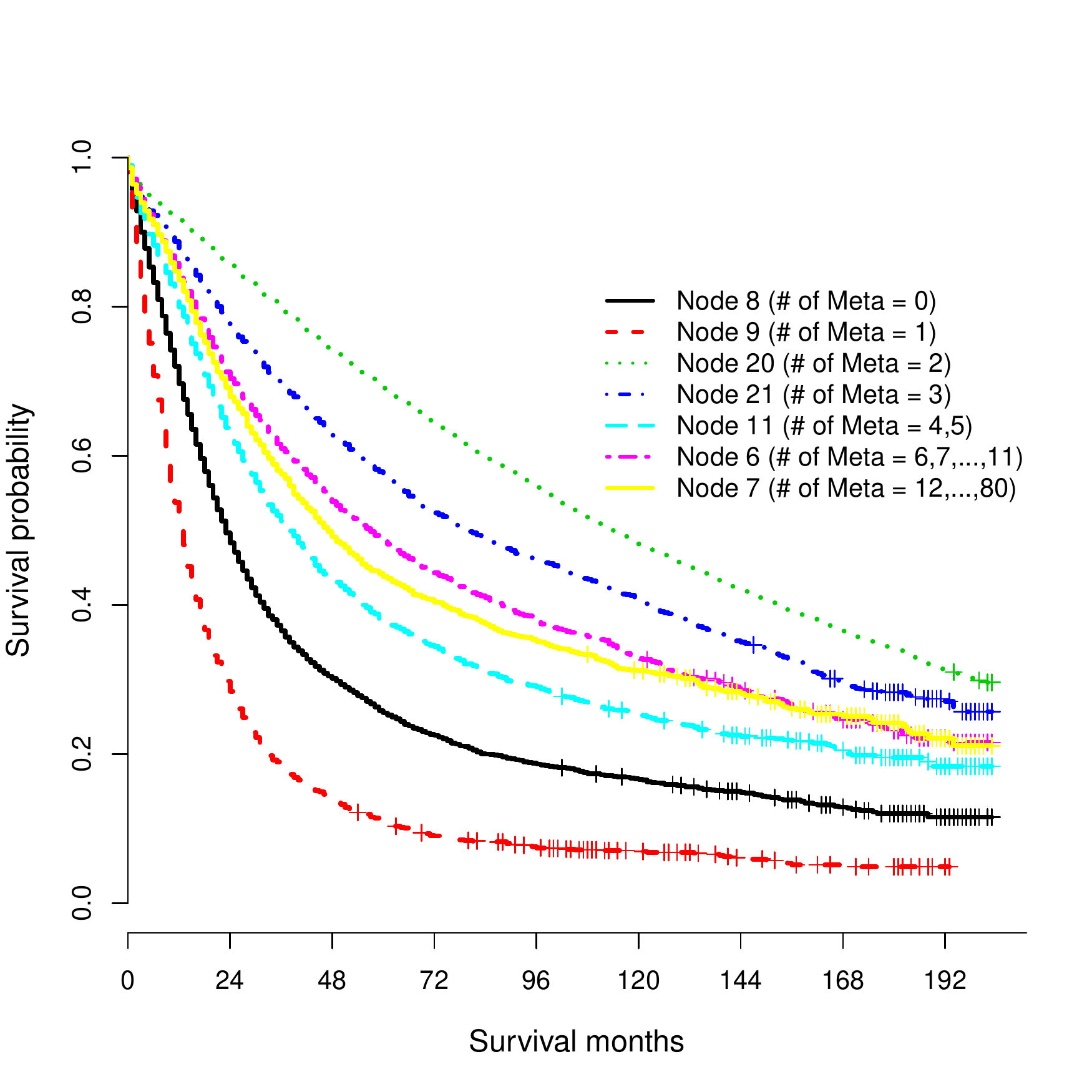}
	\caption{Kaplan-Meier survival curves of the subgroups selected by the log-rank survival tree for the colorectal cancer data.}
	\label{colon:km1}
  \end{center}
\end{figure}

For illustration, we consider the data regarding colorectal cancer \citep{SEER} from the Surveillance Epidemiology and End Results (SEER), which can be obtained from the SEER website (\url{http://seer.cancer.gov}). 
The number of metastatic LNs acts as a prognostic factor to obtain several heterogeneous subgroups with different levels of survival. 
For analysis, we selected 65,186 cases with more than 12 examined LNs because the examination of more than 12 LNs is accepted for proper evaluation of the prognosis of patients with colorectal cancers \citep{Otchy2004}. 
Figure \ref{colon:lrtree} shows a tree-diagram for the colorectal cancer data by the tree-based method used in \cite{hong2007}. 
The Kaplan-Meier survival curves for the resulting subgroups are also displayed in Figure \ref{colon:km1}. 
This indicates that survivals of some subgroups differ insignificantly or their differences are not equal-spaced. 

We propose an algorithm for overcoming these limitations and introduce a convenient software program in this paper.
Our algorithm evaluates multi-way split points simultaneously and finds an optimal set of cutpoints for a prognostic factor. 
In addition, an optimal number of cutpoints is selected by a permutation test. 
The algorithm was implemented into an \proglang{R} package \pkg{kaps}, which can be used conveniently and freely via the Comprehensive R Archive Network (CRAN, \url{http://cran.r-project.org/package=kaps}).

The rest of the paper is organized as follows. 
In Section \ref{method}, we propose novel staging algorithm called $K$-adaptive partitioning algorithm. 
In Section \ref{sec:simul}, porposed algorithm is compared with two recursive partitioning techniques through a simulation study. 
In Section \ref{exam}, the algorithm is applied to the colorectal cancer data from the SEER database. 
In Section \ref{con}, our concluding remarks are provided.
In the Appendix, an \proglang{R} package \pkg{kaps} \cite{kaps} is described and illustrated with a simple example.

\section{Proposed method} \label{method}
In this section, we describe and summarize our proposed algorithm for finding the best split set of cutpoints on a prognostic factor and for selecting the optimal number of subgroups or cutpoints in survival data. 
We call the algorithm $K$-Adaptive Partitioning for Survival data, or KAPS for short.

\subsection{Finding the best split set}
Let $T_{i}$ be a survival time, $C_i$ a censoring status, and $X_{i}$ be an ordered covariate for the $i^{th}$ observation.
We observe the triples $(Y_i, \delta_i, X_i)$ and define
\[
Y_i = \min(T_i, C_i) \; \mbox{ and } \; \delta_i = I(T_i \leq C_i),
\]
which represent the observed response variable and the censoring indicator, respectively.

Our aim is to divide the whole data $\mathfrak{D}$ into $K$ heterogeneous subgroups $\mathfrak{D}_1, \mathfrak{D}_2, \ldots, \mathfrak{D}_K$ based on the information of $X$.
All the heterogeneous subgroups should differ significantly in survival.
Rather than having both extremely poor and well separated subgroups, it is more useful to have only fairly well separated subgoups.
In other words, all the subgroups need to show greater pairwise differences than a certain criterion.
To achieve this purpose, our algorithm is constructed in the following manner.
Suppose $X$ consists of many unique values for possible splitting.
A split set (denoted by $s$) consisting of one or more cutpoints on $X$ divides the data $\mathfrak{D}$ into two or more subgroups.
That is, a split set with $(K-1)$ cutpoints generates $K$ disjoint subgroups.
There exists a number of possible split sets (denoted by $S$) because there are a number of combinations of different cutpoints on $X$.

To compare the subgroups in terms of survival, we can utilize $\chi^2$ statistics as test statistics from the log-rank or Gehan-Wilcoxon tests \cite{Peto1972}.
Let $\chi_{1}^2$ be the $\chi^2$ statistic with one degree of freedom (df) for comparing the $g^{th}$ and $h^{th}$ of $K$ subgroups created by a split set $s_K$ when $K$ is given.
For a split set $s_K$ of $\mathfrak{D}$ into $\mathfrak{D}_{1}, \mathfrak{D}_{2}, \ldots, \mathfrak{D}_{K}$, the test statistic for a measure of deviance can be defined as
\begin{equation}
   T_{1}(s_{K}) = \min_{1 \leq g < h \leq K} \chi_{1}^{2}  \; \mbox{for} \; s_{K} \in S_{K},
\end{equation}
where $S_{K}$ is a collection of split sets $s_{K}$ generating $K$ disjoint subgroups.
By this, we find the worst pair with the smallest test statistic out of the ($K-1$) adjacent pairs of $K$ subgroups constructed by $s_{K}$.
Then, take $s_{K}^{*}$ as the best split set such that
\begin{equation}
    T_{1}^{*}(s_{K}^{*}) = \max_{s_{K} \in S_{K}} T_{1}(s_{K}).
\end{equation}
The best split $s_{K}^{*}$ is a set of $(K-1)$ cutpoints which clearly separate the data $\mathfrak{D}$ into $K$ disjoint subsets of the data: $\mathfrak{D}_{1}, \mathfrak{D}_{2}, \ldots, \mathfrak{D}_{K}$.
The worst pair of the $K$ subsets should show significant differences in survival.
The overall performance can be evaluated by the overall test statistic $T_{K-1}^{*}(s_{K}^{*}) = \chi_{K-1}^2 $ statistic for comparing all $K$ subgroups from $s_{K}^{*}$.
When $K = 2$, the overall test statistic is the same as (2).
When two or more split sets have the maximum of the minimum pairwise statistics, they can be compared by their overall test statistics.
The algorithm is summarized as follows.\\

{\bf Algorithm 1.} Finding the best split set for given $K$
\begin{itemize}
\item[Step] 1: Compute chi-squared test statistics $\chi_{1}^2$ for all possible pairs, $g$ and $h$, of $K$ subgroups by $s_{K}$, where $1 \leq g < h \leq K$ and $s_{K}$ is a split set of ($K-1$) cutpoints generating $K$ disjoint subgroups.

\item[Step] 2: Obtain the minimum pairwise statistic $T_{1}(s_K)$ by minimizing $\chi_{1}^2$ for all possible pairs, $i.e.$,
$
 T_{1}(s_{K}) = \min_{1 \leq g < h \leq K} \chi_{1}^{2}  \; \mbox{for} \; s_{K} \in S_{K},
$
 where $S_{K}$ is a collection of split sets $s_{K}$ generating $K$ disjoint subsets of the data.

\item[Step] 3: Repeat Steps 1 and 2 for all possible split sets $S_{K}$.

\item[Step] 4: Take the best split set $s_{K}^{*}$ such that $T_{1}^{*}(s_{K}^{*}) = \max_{s_{K} \in S_{K}} T_{1}(s_{K})$. When two or more split sets have the maximum $T_{1}^{*}$ of the minimum pairwise statistics, choose the best split set with the largest overall statistic  $T_{K-1}^{*}$.
\end{itemize}

\subsection{Selecting the optimal number of subgroups}
One of the important issues is to determine a reasonable number of subgroups, \textit{i.e.} the selection of an optimal $K$.
The binary tree-based approaches \cite{lausen1994classification, hong2007, ishwaran2009novel, baneshi2011dichotomisation} find optimal binary splits recursively, and then determine their tree sizes using certain criteria. 
As described in Section 2.1, we find an optimal multi-way split at a time for the given number of subgroups. 
In addition, we need to choose only one of a possible number of subgroups. 
Prior information in each field may be useful. For a data-driven objective choice, we here suggest a statistical procedure to choose an optimal number of subgroups.

Let $s_{K}^{*}$ and $T_{1}^{*} (s_{K}^{*})$ be the best split set and the minimum pairwise statistic using the raw data for each $K$.
The data can be reconstructed by matching their labels after permuting the labels of $X$ with retaining the labels of $(Y, \delta)$.
The survival time $Y$ is independent of the covariate $X$ in the reconstructed data, which is called the permuted data.
When the permuted data are allocated into each subgroup by $s_{K}^{*}$, there should be no significant differences in survival among the subgroups.
The repetition of this procedure generates the null distribution of the test statistics. 
If we repeat this procedure many times ($R$ times), and then we obtain the permutation $p$-value $p_{K}$ for each $K$.
This is the ratio where the minimum pairwise statistics of the permuted data are greater than or equal to that of the raw data,  $i.e.$,
\[
p_{K} = \sum_{r=1}^{R} I( T_{1}^{(r)} (s_{K}^{*}) \geq T_{1}^{*} (s_{K}^{*}))/R, \; K = 2, 3, \ldots ,
\]
where $T_{1}^{(r)} (s_{K}^{*})$ is the $r^{th}$ repeated minimum pairwise statistic for the permuted data.
In addition, we correct the $p$-values for multiple comparison because there are ($K-1$) comparisons between two adjacent subgroups when there are $K$ subgroups.
For example, the corrected $p$-value can be obtained using Bonferroni correction, $i.e.$,  $p_{K}^{c} = {p_{K}}/{(K-1)}, K = 2, 3, \ldots$. 
Lastly, we choose the largest number to discover as many significantly different subgroups as possible, given that the corrected $p$-values are smaller than or equal to a pre-determined significance level, $e.g.$  $\alpha = 0.05$. Formally,
\begin{equation}
	\hat{K} = \max \{K |\; p_{K}^{c} \leq \alpha, K = 2, 3, \ldots\}.
\end{equation}
The algorithm is summarized as follows.
\\

{\bf Algorithm 2.} Selecting the optimal number of subgroups ($K$)
\begin{itemize}
\item[Step] 1: Find $s_{K}^{*}$ and $T_{1} (s_{K}^{*})$ with the raw data for each $K$ using Algorithm 1.

\item[Step] 2: Construct the permuted data by permuting the labels of $X$ whilst retaining the labels of $(Y, \delta)$.

\item[Step] 3: Allocate the permuted data into each subgroup by $s_{K}^{*}$.

\item[Step] 4: Obtain the minimum pairwise statistic $T_{1}^{(r)} (s_{K}^{*})$ for the permuted data.

\item[Step] 5: Repeat steps 2 to 4 $R$ times, and then obtain $T_{1}^{(1)} (s_{K}^{*}), T_{1}^{(2)} (s_{K}^{*}), \ldots, T_{1}^{(R)} (s_{K}^{*})$.

\item[Step] 6: Compute the permutation $p$-value $p_{K}$ for each $K$, $i.e.$,
$
p_{K} = \sum_{r=1}^{R} I( T_{1}^{(r)} (s_{K}^{*}) \geq T_{1} (s_{K}^{*}))/R, \; K = 2, 3, \ldots .
$

\item[Step] 7: Correct the permutation $p$-value $p_{K}$ by correcting for multiple comparisons, $e.g.$, corrected $p$-value $p_{K}^{c} = {p_{K}}/{(K-1)}, K = 2, 3, \ldots$.

\item[Step] 8: Select the largest $K$ when the corrected $p$-values are less than or equal to $\alpha$, $i.e.$,
$
\hat{K} = \max \{K |\; p_{K}^{c} \leq \alpha, K = 2, 3, \ldots\}.
$
\end{itemize}


\section{Simulation study}
\label{sec:simul}
In this section, we investigate the performance of our proposed method (\code{kaps}) with simulated data.
For comparison, we employ two recursive partitioning algorithms: survival CART \cite{leblanc1992relative} and conditional inference tree \cite{hothorn2006unbiased}.
The former is descended from the traditional CART for survival data and the latter is based on maximally selected rank statistics \cite{hothorn2003}.
They were implemented in the \proglang{R} packages \pkg{rpart} \cite{rpart} and  \pkg{party} \cite{hothorn2006unbiased}, respectively. Our algorithm was implemented in the R package \pkg{kaps} \cite{kaps}.

\subsection{Simulation setting}
To generate simulated data, we assume that survival time $T_i$ is generated from exponential distribution with a parameter $\lambda_i$ and and censoring times $C_i$ is generated from Uniform distribution with appropriate parameters.
Then we observe $Y_i = \min(T_i, C_i)$ and $\delta_i = I(T_i \leq C_i)$, where $i = 1,2,\ldots, n$. In addition, a prognostic factor $X_i$ is generated from a discrete uniform distribution with a range of 1 and 20, {\em i.e.}, DU$(1,20)$. 
We first consider the following stepwise model (SM) defining parameter $\lambda_i$ as follows.
\[
\lambda_i = \begin{cases}
 0.02, & \;\;\;\;\;\;\;\;\;   X_i  \leq 7, \\
 0.04, & \;\; 7 <   X_i  \leq 14, \\
 0.08, & 14 <  X_i,
\end{cases}
\]
This model has three different hazard rates that are distinguished by two cutpoints 7 and 14. In addition, we consider the following linear model (LM) defining parameter $\lambda_i$ as follows.
\[
	\lambda_i = 0.1 X_i.
\]
In this model, $\lambda_i$ depends on $X_i$ linearly. It follows that $Y_i$ depends on $X_i$ nonlinearly. 
This model has a number of different hazard rates. 
For each model, we generate a simulated data set of 200 observations with average censoring rates of 15\% or 30\%. 
For testing, we independently generate a test data set of sample size 200 observations. 
All the simulation experiments are repeated 100 times independently.

\subsection{Simulation results}

\begin{table}
\centering
\caption{Overall and minimum pairwise log-rank statisitics (standard errors) by each of \code{rpart}, \code{ctree}, and \code{kaps} for the stepwise model (SM) and linear model (LM) with average censoring rates (CR) of 15\% or 30\%. The minimum pairwise statistic is the smallest one among all the pairs. For SM, the log-rank statistics are provided for reference (\code{ref}) when the true cutpoints are used.}
\label{simul:tab1}
{\small
\begin{tabular}{cclrrrr}
\hline
Model & K &  \multicolumn{1}{c}{Method} & \multicolumn{2}{c}{CR = 15\%} &  \multicolumn{2}{c}{CR = 30\%} \\
 & &  & \multicolumn{1}{c}{Overall} & \multicolumn{1}{c}{Pairwise}  & \multicolumn{1}{c}{Overall} & \multicolumn{1}{c}{Pairwise} \\
\hline
    &  & \code{ref} & 48.68 (1.17) & 9.06 (0.42) & 39.84 (1.29) & 7.13 (0.37) \\
    &   & \code{rpart} & 35.39 (1.29) & 3.88 (0.43) & 27.55 (1.24) & 2.56 (0.30)\\
SM  & 3 & \code{ctree} & 38.97 (1.28) & 5.21 (0.46) & 30.90 (1.21) & 3.94 (0.38) \\
    &  & \code{kaps} & 39.69 (1.35) & 7.11 (0.47) & 31.42 (1.27) & 5.04 (0.38) \\
\hline
&   &\code{rpart} & 43.57 (1.17) & 43.57 (1.17) & 38.73 (1.03) & 38.73 (1.03) \\
& 2 &\code{ctree} & 38.94 (1.16) & 38.94 (1.16) & 36.29 (1.01) & 36.29 (1.01) \\
&   &\code{kaps} &  44.52 (1.15) & 44.52 (1.15) & 38.74 (1.11) & 38.74 (1.11) \\
\cline{2-7}
   &  &\code{rpart} & 48.07 (1.47) & 6.87 (0.50) & 43.21 (1.11) & 6.32 (0.44)\\
LM & 3  &\code{ctree} &55.64 (1.39) & 12.94 (1.14) & 47.99 (1.40) & 8.33 (0.64) \\
   &    &\code{kaps} & 54.96 (1.39) & 13.83 (0.63) & 47.95 (1.36) & 11.33 (0.59)\\
\cline{2-7}
&   &\code{rpart} & 59.60 (1.59) & 2.59 (0.26) & 53.01 (1.33) & 2.30 (0.22) \\
& 4 &\code{ctree} & 59.82 (1.35) & 3.17 (0.37) & 52.05 (1.34) & 2.10 (0.21) \\
&   &\code{kaps} & 61.27 (1.39) & 3.22 (0.24) & 53.48 (1.34) & 2.70 (0.23) \\
\hline
\end{tabular}
}
\end{table}

\begin{figure}	
        \centering
        \begin{subfigure}[b]{0.4\textwidth}
               \includegraphics[width=\textwidth]{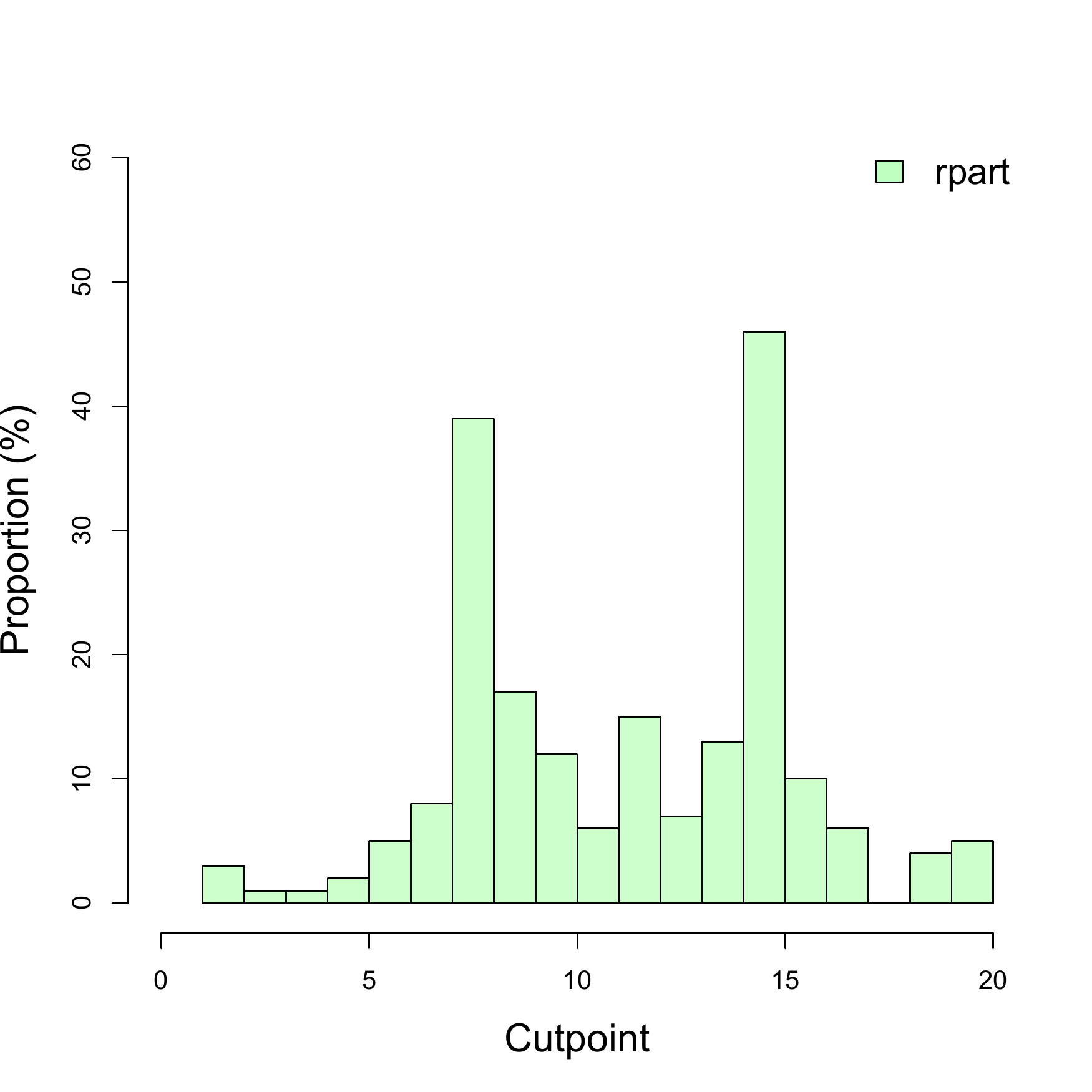}
	      \caption{\code{rpart}}
        \end{subfigure}%
	\quad
        \begin{subfigure}[b]{0.4\textwidth}
                \includegraphics[width=\textwidth]{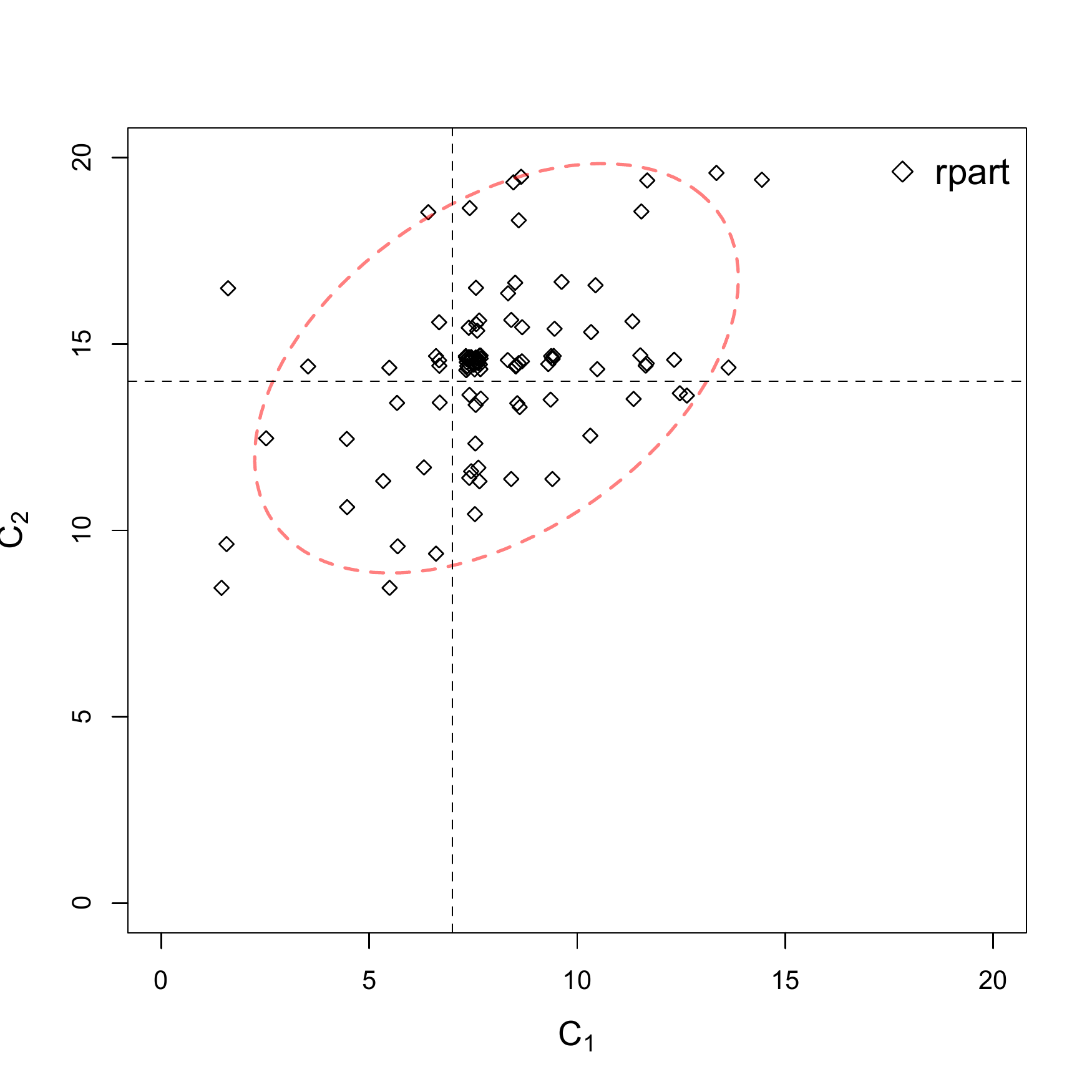}
	      \caption{\code{rpart}}
        \end{subfigure}
	\\
        \begin{subfigure}[b]{0.4\textwidth}
                \includegraphics[width=\textwidth]{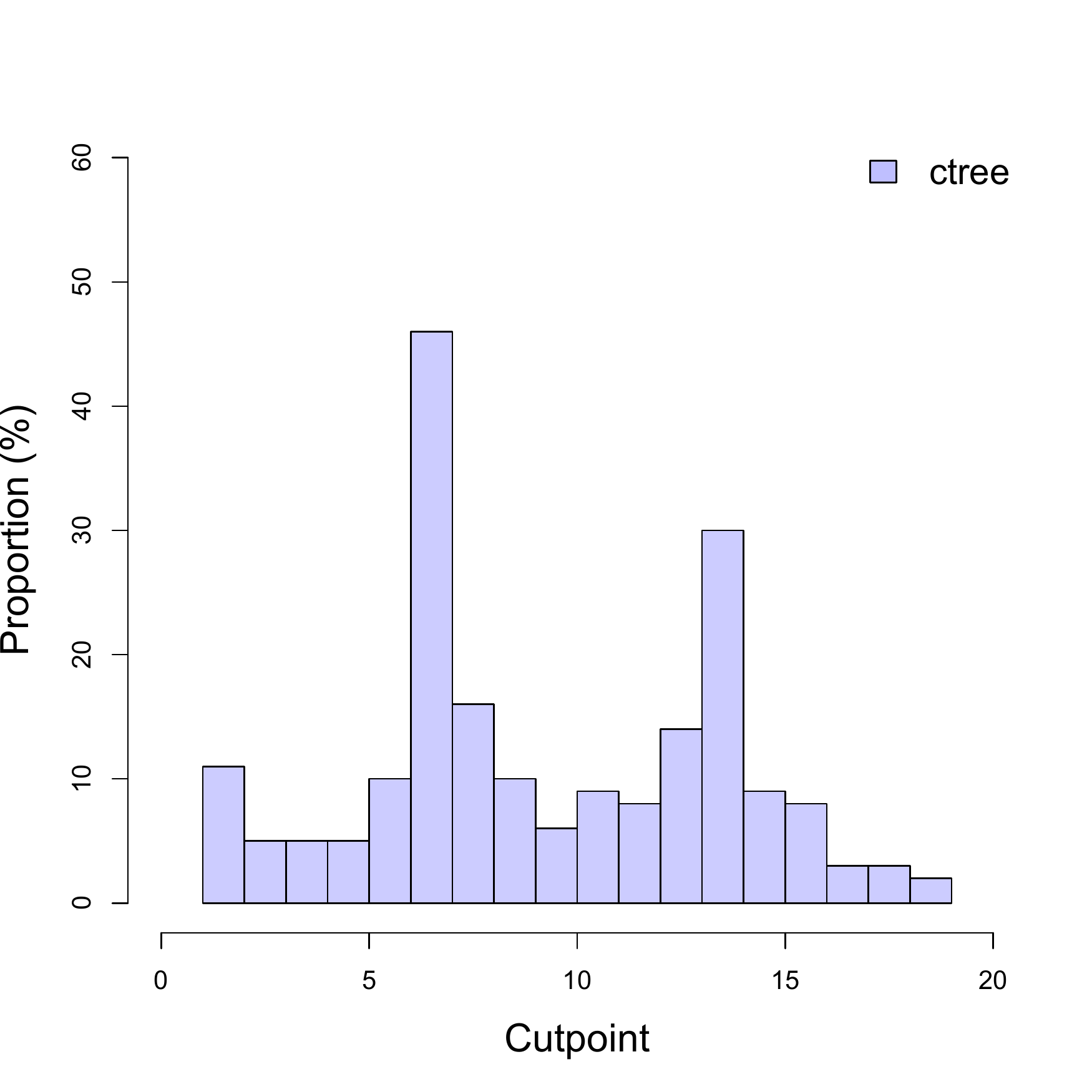}
	      \caption{\code{ctree}}
        \end{subfigure}%
	\quad
        \begin{subfigure}[b]{0.4\textwidth}
                \includegraphics[width=\textwidth]{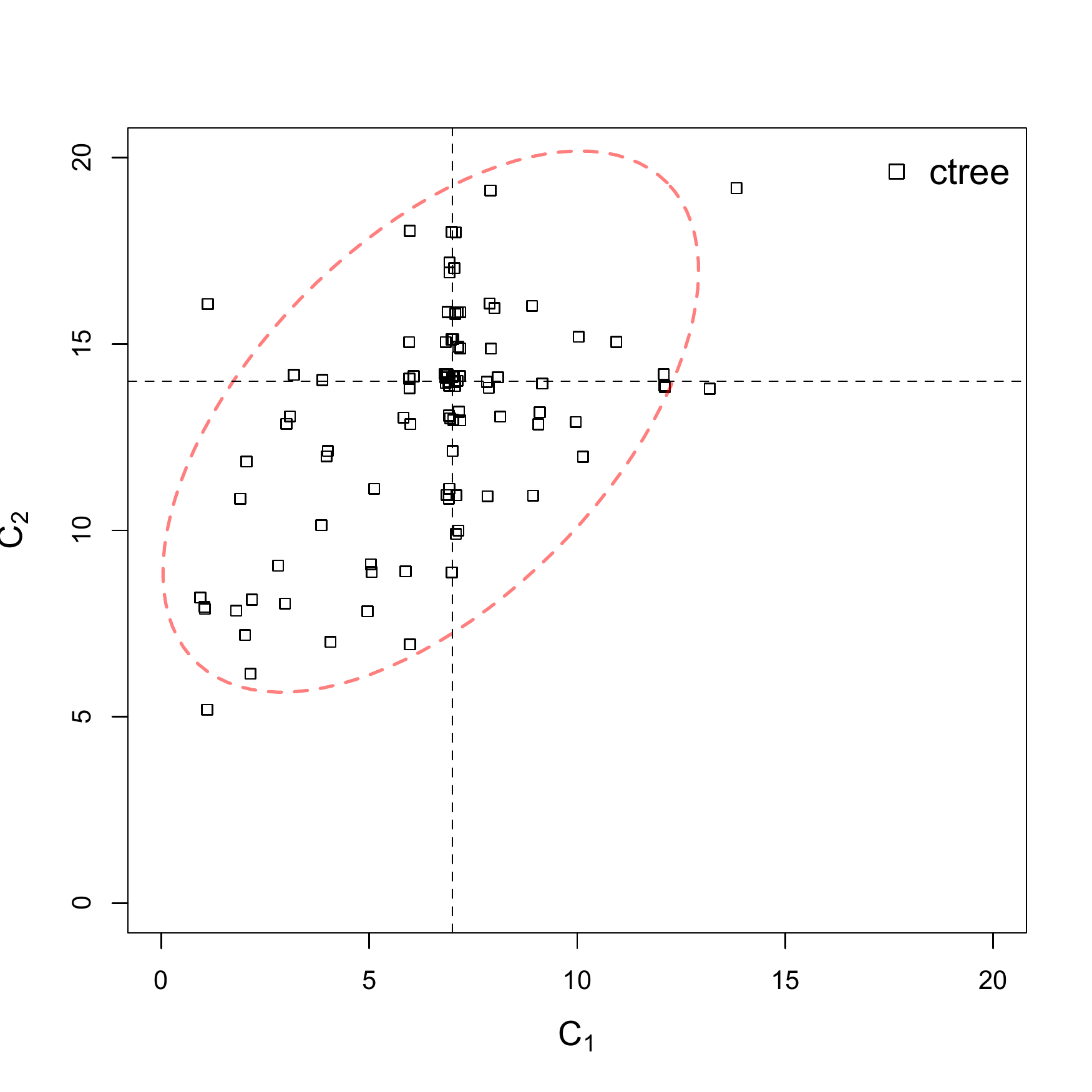}
	      \caption{\code{ctree}}
        \end{subfigure}
	\\
        \begin{subfigure}[b]{0.4\textwidth}
                \includegraphics[width=\textwidth]{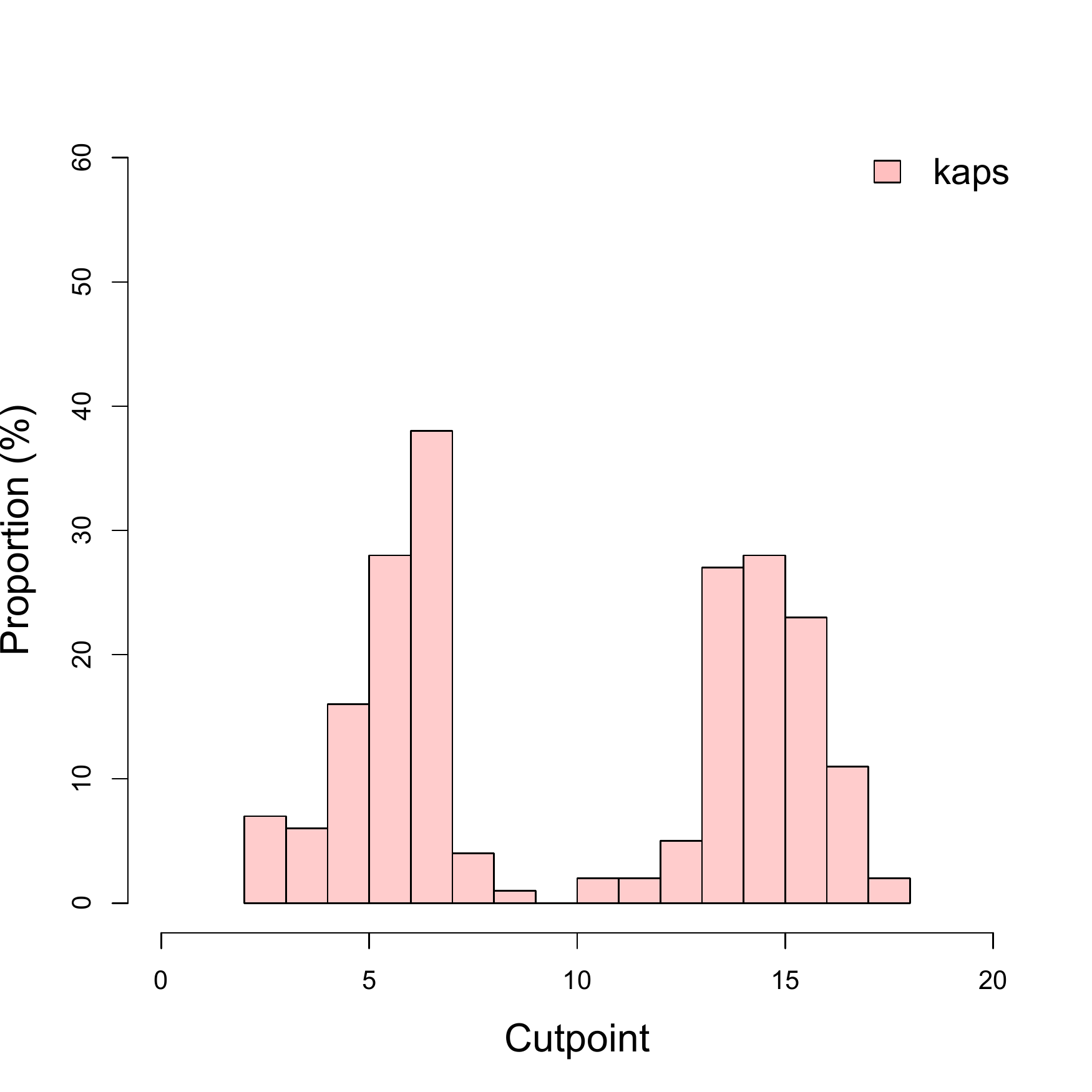}
	      \caption{\code{kaps}}
        \end{subfigure}%
	\quad
        \begin{subfigure}[b]{0.4\textwidth}
                \includegraphics[width=\textwidth]{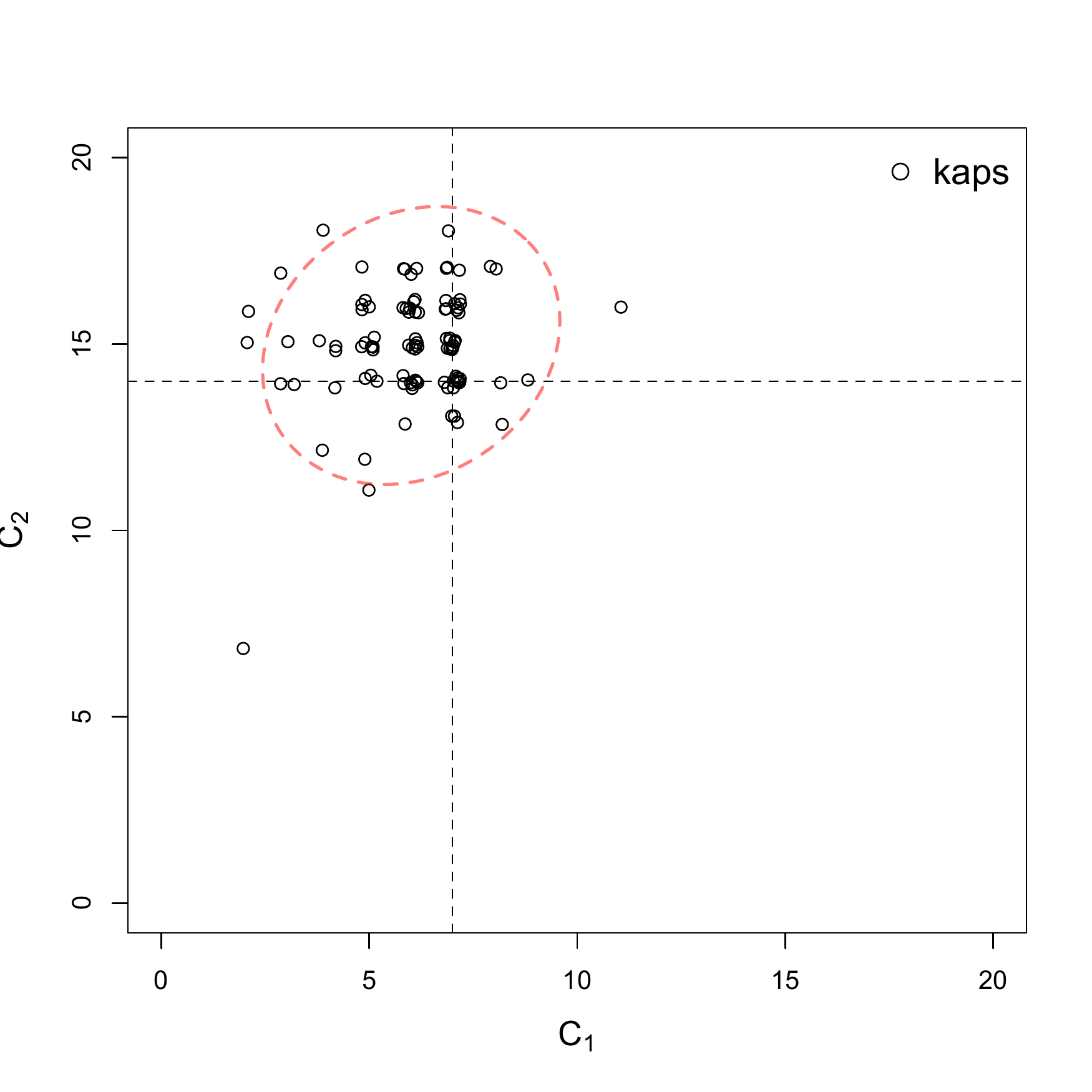}
	      \caption{\code{kaps}}
        \end{subfigure}
        \caption{Cutpoint selection for the stepwise model (SM) with average censoring rate 30\%. The histograms and the scatterplots of two selected cutpoints, $C_1$ and $C_2$, for each method are displayed in the left and right columns, respectively. The 95\% confidence ellipse is superimposed on each scatterplot.}
	\label{simul:case2:fig}
\end{figure}

\begin{figure}
        \centering
        \begin{subfigure}[b]{0.45\textwidth}
                \includegraphics[width=\textwidth]{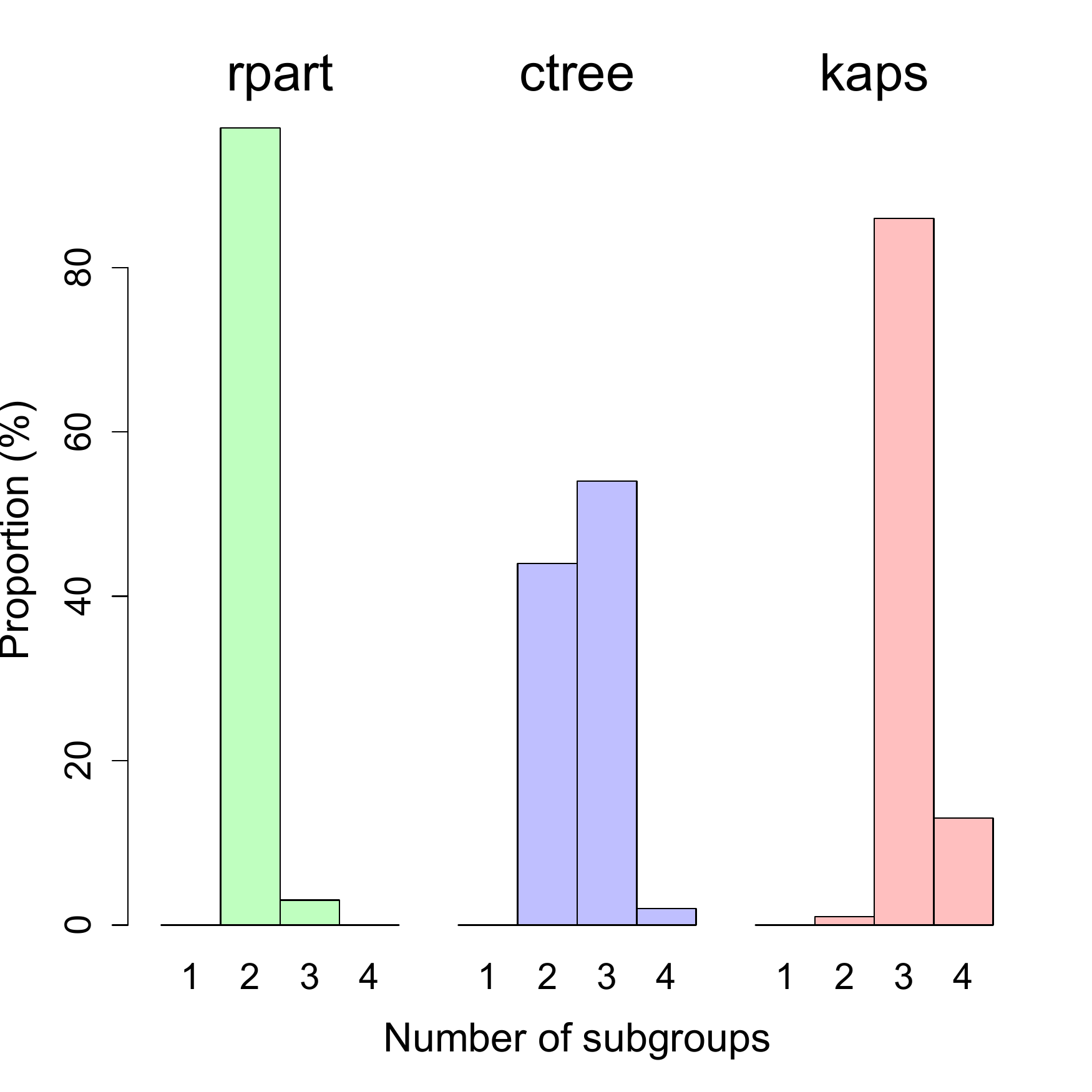}
	      \caption{SM model with CR = 15\% }
        \end{subfigure}%
	\quad
        \begin{subfigure}[b]{0.45\textwidth}
                \includegraphics[width=\textwidth]{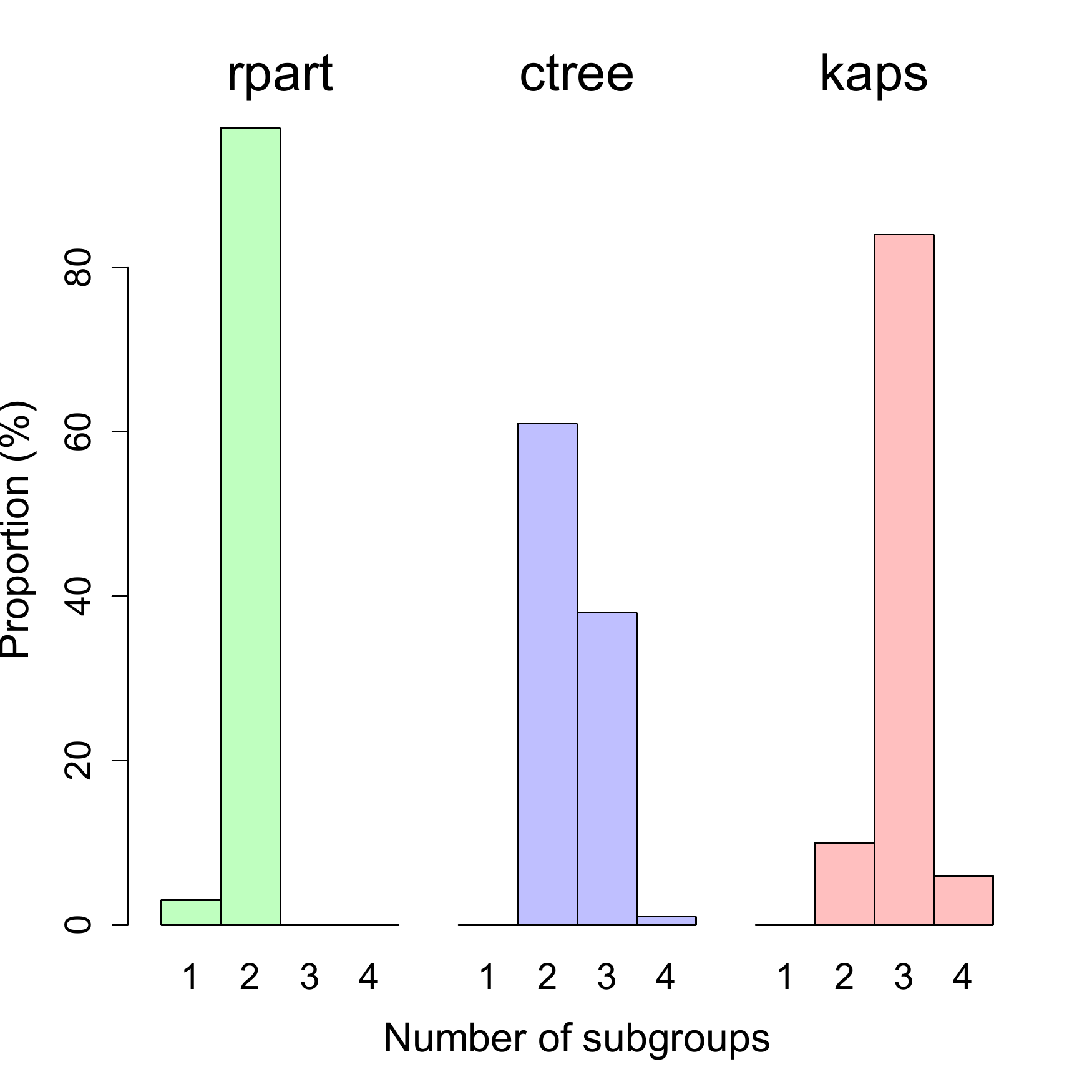}
	      \caption{SM model with CR = 30\% }
        \end{subfigure}
	\\
        \begin{subfigure}[b]{0.45\textwidth}
                \includegraphics[width=\textwidth]{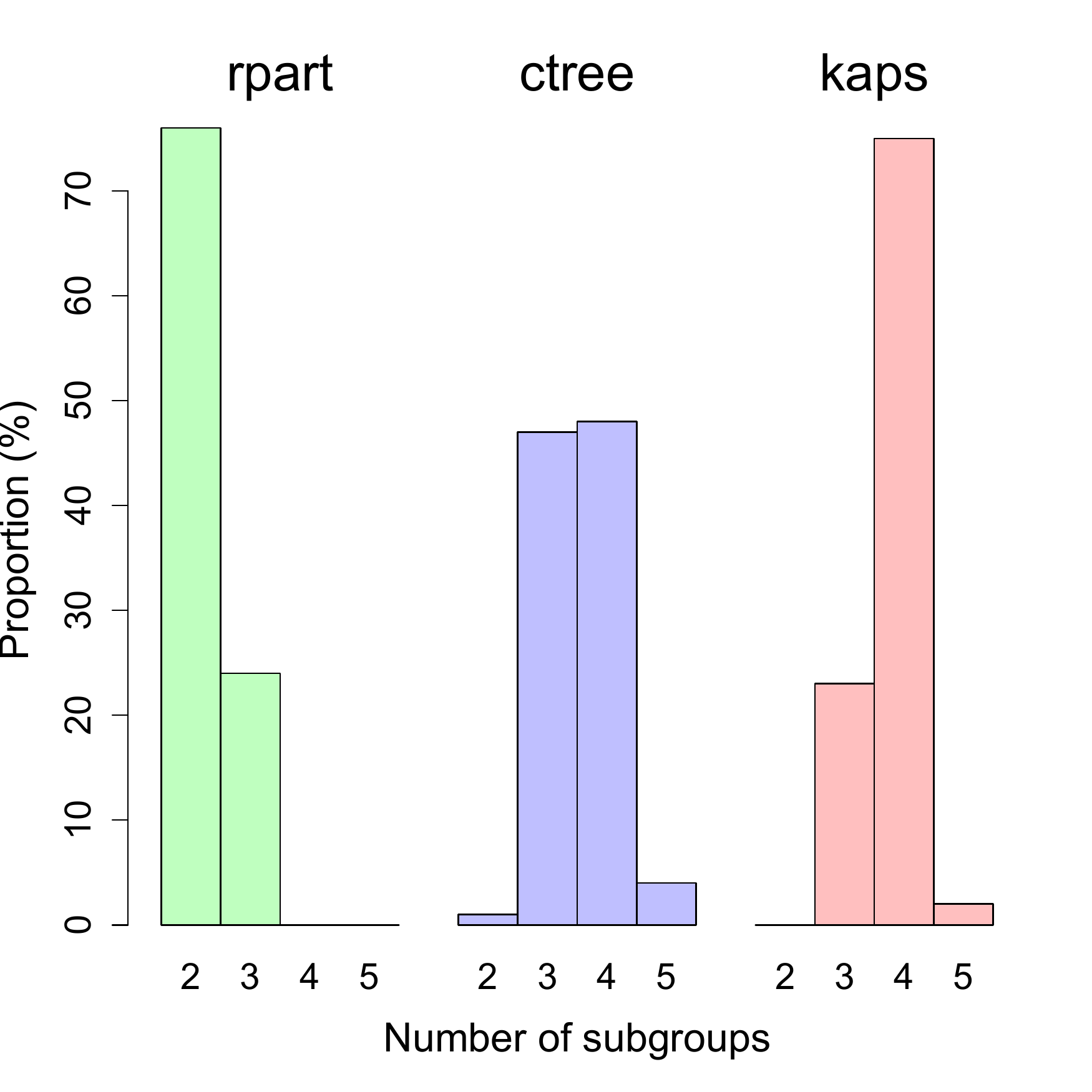}
	      \caption{LM model with CR = 15\% }
        \end{subfigure}%
	\quad
        \begin{subfigure}[b]{0.45\textwidth}
                 \includegraphics[width=\textwidth]{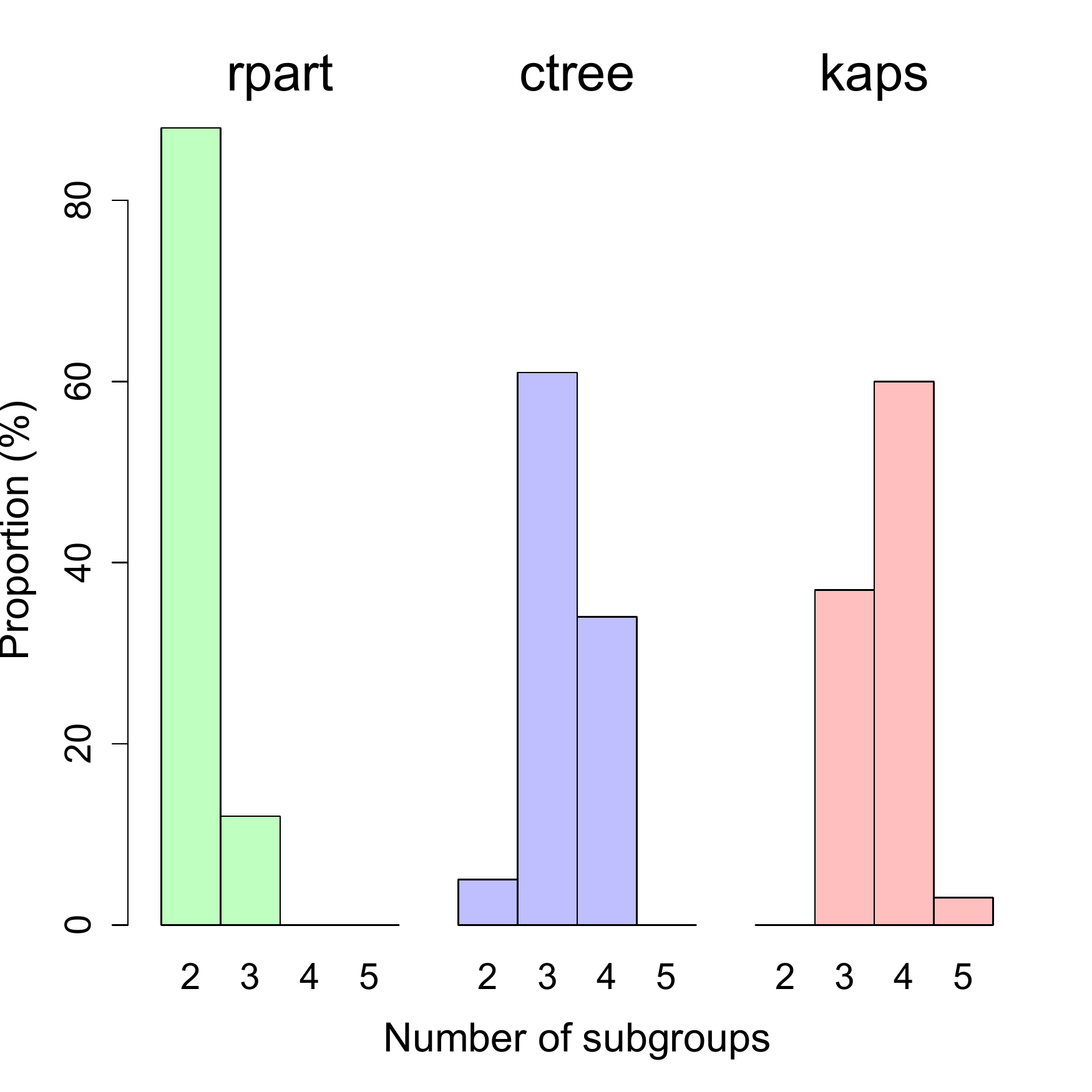}
 	      \caption{LM model with CR = 30\% }
         \end{subfigure}
        \caption{Proportions of the selected numbers of subgroups by rpart, ctree, and \code{kaps} for the stepwise model (SM) and linear model (LM) with averaged censoring rate (CR) 15\% or 30\%}
	\label{simul:subgroup}
\end{figure}

We first study whether the selection of cutpoints is correct when the number of subgroups $K$ is specified. 
In addition, we investigate whether the cutpoint selection affects the partition performance, which is measured by the overall log-rank statistic and the minimum pairwise log-rank statistic. 
The overall log-rank statistic is for testing the differences of all the subgroups and the minimum pairwise log-rank statistic is the smallest from all the pairs.

For the SM, it is reasonable to select cutpoints 7 and 14 because the hazard rates are distinguished by these two cutpoints. 
On the other hand, it is not clear which points should be selected for the LM. 
Thus, we investigate the frequencies of selected cutpoints by each of \code{rpart}, \code{ctree}, and \code{kaps} for the SM. 
As seen in the histograms in Figure \ref{simul:case2:fig}, \code{kaps} often selects points \textit{around} 7 and 14, while \code{rpart} and \code{ctree} often tend to select the points \textit{between} 7 and 14. 
The scatterplots in Figure \ref{simul:case2:fig} show the distributions of the selected cutpoints in two-dimensional space where each axis indicates each cutpoint. 
The cutpoints of \code{kaps} are mostly distributed in a smaller ellipse (almost circular), while those of \code{rpart} and \code{ctree} are distributed in larger ellipses.
Therefore, we can say that \code{kaps} selects the true cutpoints better than \code{rpart} and \code{ctree}.

For SM, the two true cutpoints 7 and 14 are known. Thus, when the true cutpoints are used for partitioning, the overall statistics are 48.68 and 39.84 and the minimum pairwise statistics are 9.06 and 7.13 when CR are 15\% and 30\%, respectively. 
It can be shown that \code{kaps} has the largest overall and minimum pairwise statistics from the three methods, while \code{rpart} has the smallest value of these statistics.
Therefore, the correct selection of cutpoints leads to an improved performance in partitioning.
For LM, the true cutpoints are unknown. 
Moreover, it is not known how many subgroups will be best. 
Thus, we assume 2, 3, and 4 subgroups ($K = 2, 3, 4$) because these would be useful in practice. 
When $K = 2$, the overall statistics are the same as the pairwise statistics because there are only two subgroups. 
The results show that \code{kaps} is slightly better and \code{ctree} is slightly worse than the others, although all the methods perform well. 
When $K = 3$, all the methods lead to significant differences between all pairs of subgroups. 
However, \code{kaps} performs the best, while \code{rpart} is the worst. 
When $K = 4$, none of the methods find a significant difference for the worst pair although \code{kaps} is slightly better. 
This implies that it is reasonable to have three subgroups ($K = 3$) in this case.

We next explore how many subgroups are selected by each method. For each method, the default option was used and the minimum sample size in each subgroup was 10\% of the data.
Figure \ref{simul:subgroup} displays the histograms of the subgroups selected by each method for SM and LM with CR of 15\% or 30\%. 
For SM, \code{rpart} tends to identify two subgroups and \code{ctree} identifies two or three subgroups. 
In contrast, \code{kaps} most often identifies three subgroups. For LM, \code{rpart} tends to identify two subgroups and \code{ctree} identifies three or four subgroups. 
On the other hand, \code{kaps} identifies three or four, but mostly four subgroups. 
This implies that \code{rpart} identifies a smaller number of subgroups  while \code{kaps} does a larger number.

\section{Example}\label{exam}

\begin{table}
\centering
\caption{Numbers of metastasis lymph nodes for the staging systems by \code{AJCC}, \code{kaps}, \code{ctree}, and \code{rpart}. Minimum pairwise statistics and their corresponding pairs of subgroups are given at the bottom.}
\label{colon:staging}
{\small
\begin{tabular}{crrrr}
\hline
 Subgroup   & \multicolumn{1}{c}{\code{AJCC}}  & \multicolumn{1}{c}{\code{rpart}} & \multicolumn{1}{c}{\code{ctree}} & \multicolumn{1}{c}{\code{kaps}} \\
\hline
 Subgroup 1 & 0           & 0           & 0       & 0         \\
 Subgroup 2 & 1           & 1,2,3       & 1       & 1         \\
 Subgroup 3 & 2,3         & 4 $\sim$ 10 & 2,3     & 2,3       \\
 Subgroup 4 & 4,5,6       & $\geq$ 11   & 4,5     & 4,5,6     \\
 Subgroup 5 & $\geq$7     & \textemdash & 6,7,8   & 7,8,9,10  \\
 Subgroup 6 & \textemdash & \textemdash & $\geq$9 & $\geq$ 11 \\
\hline
\multicolumn{1}{l}{Min. pairwise statistic}& 131.23 & 932.30 & 78.35 & 131.23  \\
\multicolumn{1}{l}{Corresponding pair} & (2, 3) & (3, 4) & (3, 4) & (2, 3) \\
\hline
\end{tabular}
}
\end{table}

\begin{figure}[!t]
        \centering

        \begin{subfigure}[b]{0.45\textwidth}
                \includegraphics[width=\textwidth]{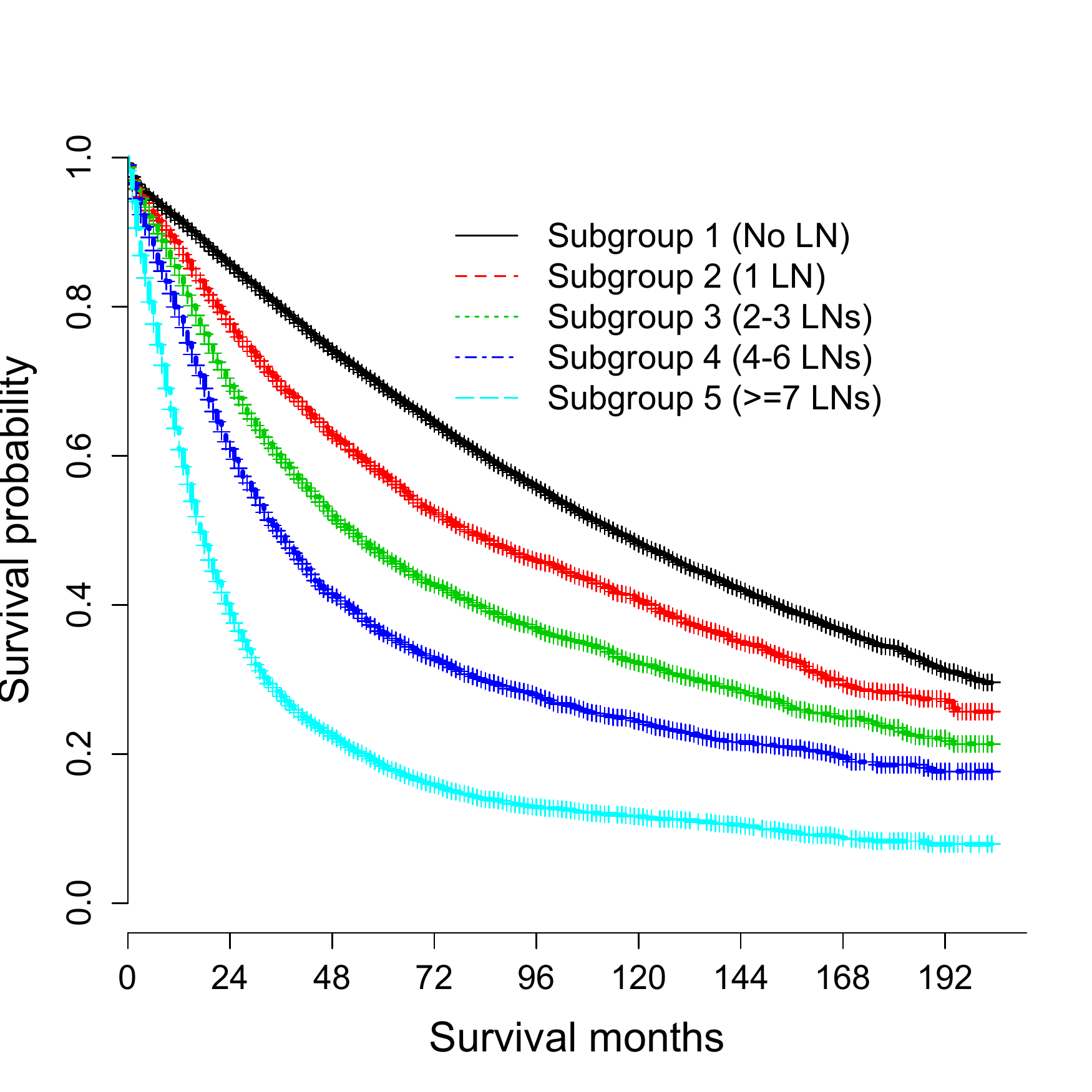}
	      \caption{\code{AJCC}}
        \end{subfigure}%
	\quad
        \begin{subfigure}[b]{0.45\textwidth}
                \includegraphics[width=\textwidth]{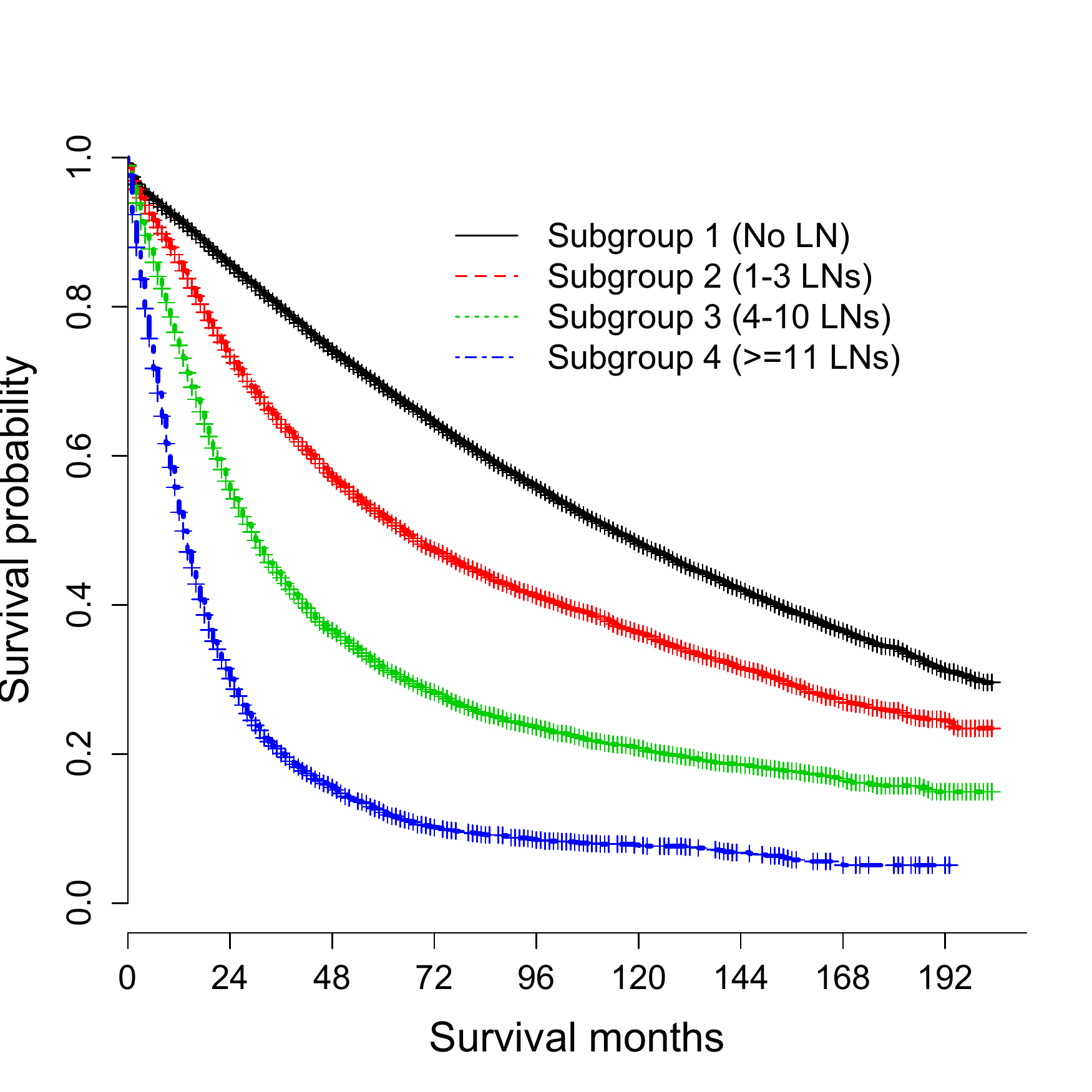}
	      \caption{\code{rpart}}
        \end{subfigure}
	\\
        \begin{subfigure}[b]{0.45\textwidth}
                \includegraphics[width=\textwidth]{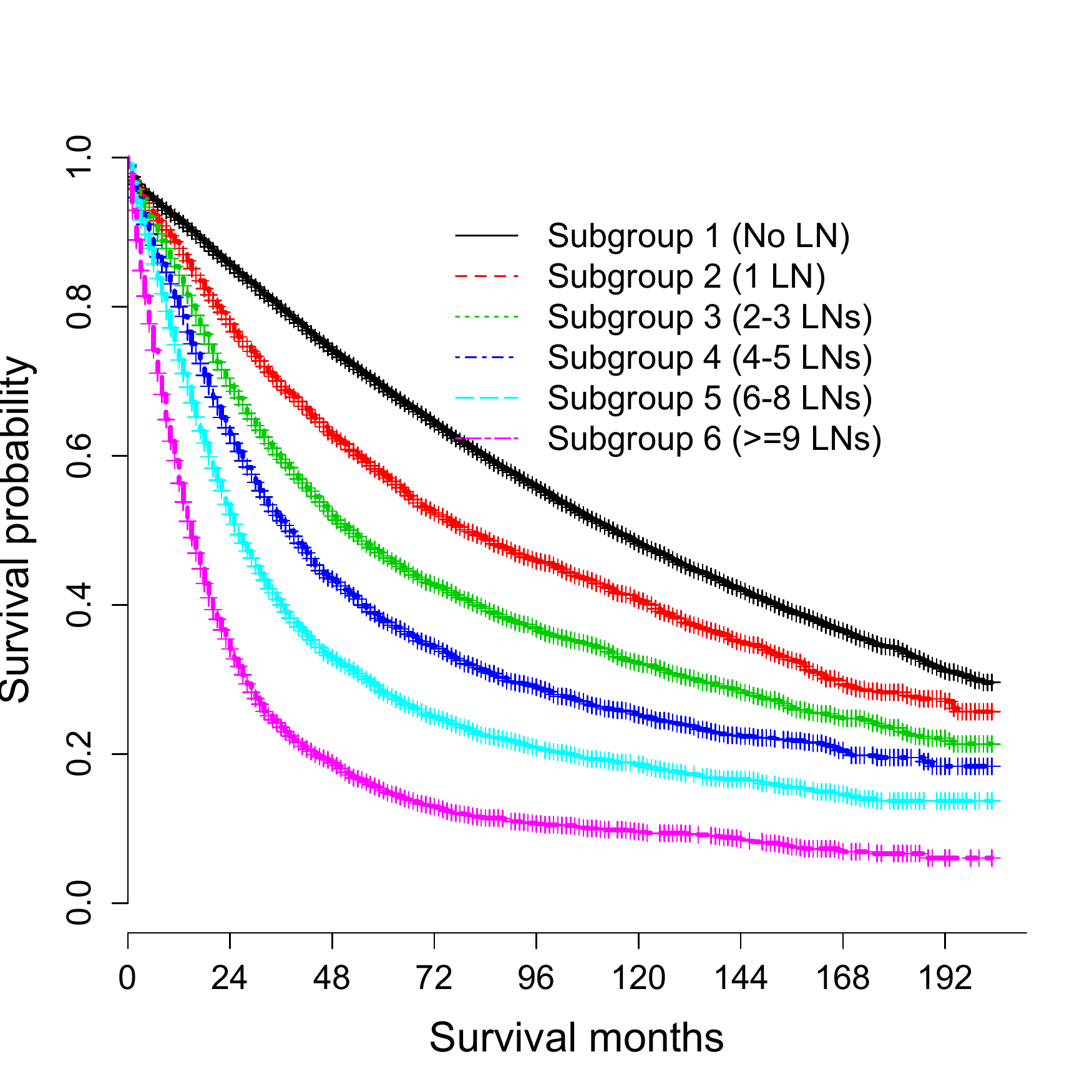}
	      \caption{\code{ctree}}
        \end{subfigure}%
	\quad
        \begin{subfigure}[b]{0.45\textwidth}
                \includegraphics[width=\textwidth]{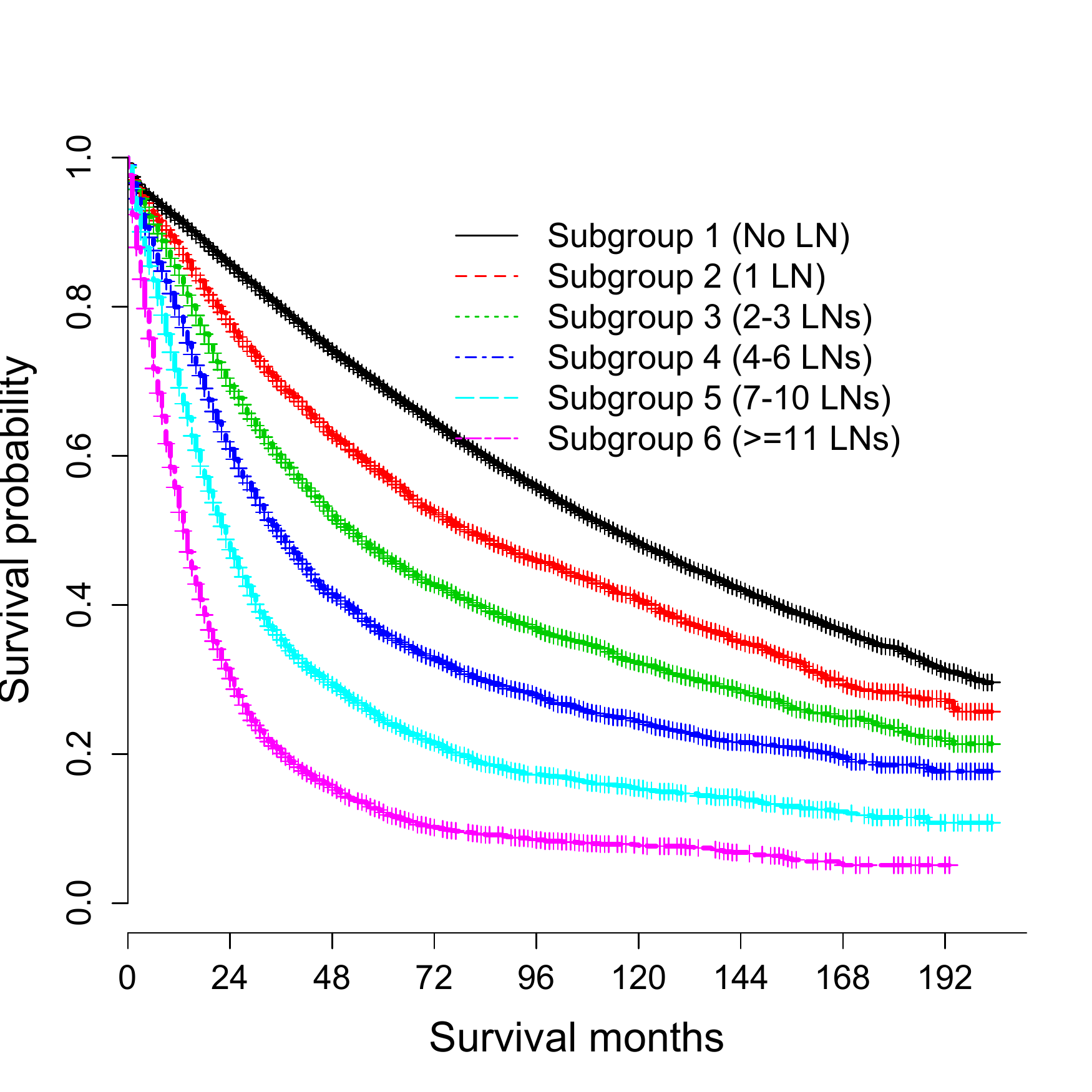}
	      \caption{\code{kaps}}
        \end{subfigure}

        \caption{Kaplan-Meier survival curves for the subgroups identified by \code{AJCC}, \code{rpart}, \code{ctree}, and \code{kaps}}
	\label{colon:fig}
\end{figure}
We apply our proposed method to the colorectal cancer data from the Surveillance Epidemiology and End Results (SEER) database (\url{http://seer.cancer.gov}). 
The SEER data includes information about a variety of cancers and has been collected from various locations and sources in the US since 1973 and it is continually expanded to cover more areas and demographics. 
It includes incidence and population data associated with age, gender, race, year of diagnosis, and geographic areas. 
We here utilized the data consisting of patients with colorectal cancer, which were used to develop a new cancer staging system. 
We use the number of metastatic lymph nodes (LNs) as an ordered prognostic factor, which was used for the N classification of the current TNM staging system of the American Joint Committee of Cancer (\code{AJCC}). For analysis, 65,186 cases were selected with 12 or more examined LNs because this many LNs need to be examined for evaluating the prognosis of colorectal cancer patients \cite{Otchy2004}.

Table \ref{colon:staging} shows the numbers of metastatic LNs for the stages discovered by \code{rpart}, \code{ctree}, and \code{kaps}, including the N classification of the current TNM staging system of American Joint Committee of Cancer (\code{AJCC}). 
The minimum pairwise log-rank statistics and their corresponding pairs of subgroups are given at the bottom of this table. 
\code{AJCC} consists of 5 subgroups and the worst pair of subgroups is (2, 3) with a minimum pairwise log-rank statistic of 131.23. 
That is, two subgroups 2 and 3 are \{LNs = 1\} and \{LNs = 2 or 3\}. 
\code{rpart} has the largest minimum pairwise statistic, but it discovers only 4 subgroups. 
In contrast, \code{ctree} and \code{kaps} identify one more subgroup than \code{AJCC}.  
Our \code{kaps} has a smaller minimum pairwise statistic than \code{ctree}. 
Thus, we can say that \code{kaps} performs better. 
The survival curves for the subgroups are shown in Figure \ref{colon:fig}. 
\code{rpart} shows well-separated curves, but they are only 4 subgroups. 
Our \code{kaps} consists of 6 subgroups, all of which are shown to be fairly well-separated in Figure \ref{colon:fig}.

\section{Conclusion} \label{con}
In this paper, we have proposed a multi-way partitioning algorithm for censored survival data. It divides the data into $K$ heterogeneous subgroups based on the information of a prognostic factor.
The resulting subgroups show significant differences in survival. 
Rather than a mixture of extremely poorly and well-separated subgroups, our developed algorithm aims to generate only fairly well-separated subgroups even though there is no extremely well-separated subgroup. 
For this purpose, we identify a multi-way partition which maximizes the minimum of the pairwise test statistics among subgroups. 
The partition consists of two or more cutpoints, whose number is determined by a permutation test.

Our developed algorithm is compared with two binary recursive partitioning algorithms, which are widely used in \proglang{R}. 
The simulation study implies that our algorithm outperforms the others. 
In addition, its usefulness was demonstrated using a real colorectal cancer data set from the SEER database. 
We have implemented our algorithm in an \proglang{R} package \pkg{kaps}, which is convenient to use and freely available in \proglang{R} via the Comprehensive R Archive Network (CRAN, \url{http://cran.r-project.org/package=kaps}).

\section*{Acknowledgement}
This research was supported by the Basic Science Research Program through the National Research Foundation of Korea (NRF) funded by the Ministry of Education, Science and Technology (2010-0007936). It was also supported by the Asan Institute for Life Sciences, Seoul, Korea (2013-554).


\newpage
\section*{Appendix}

The algorithm described in this paper was implemented into an \proglang{R} package \pkg{kaps} \cite{kaps} which is available at the Comprehensive R Archive Network (CRAN, \url{http://cran.r-project.org/package=kaps}).
In the Appendix, we illustrate the use of the algorithm with a simple example.

\subsection*{Overview}
A package \pkg{kaps} was written in \proglang{R} language \citep{Rlang} which allows clean interface implementation and great extension. 
The package depends on \pkg{methods}, \pkg{survival} \citep{survival}, \pkg{Formula} \citep{Formula} and \pkg{coin} \citep{zeileis2008implementing} packages. 
The package \pkg{Formula} is utilized to handle  multiple parts on the right-hand side of the \code{formula} object for convenient use. 
The package \pkg{coin} is used for the permutation test for the selection of optimal number of subgroups. 
In addition, the packages \pkg{locfit} \citep{locfit}, \pkg{foreach} \citep{foreach} and \pkg{doMC} \citep{doMC} are suggested to give fancy visualization and minimize computational cost, respectively. 

\subsection*{Main Function}
The $K$-adaptive partitioning algorithm can be conducted by a function \code{kaps()}.  
Usage and input arguments for \code{kaps()} are as follows. 
The type of the arguments is given in brackets.
\begin{Schunk}
\begin{Sinput}
kaps(formula, data, K = 2:4, mindat, type = c("perm", "NULL"), ...)
\end{Sinput}
\end{Schunk}
\begin{itemize}
\item \code{formula} [S4 class \code{Formula}]: a \code{Formula} object with a response variable on the left hand side of the $\sim$ operator and covariate terms on the right side. The response has to be a survival object with survival time and censoring status in the \code{Surv} function.
\item \code{data} [data.frame]: a data frame with variables used in the \code{formula}. It needs at least three variables including survival time, censoring status, and a covariate.
\item \code{K} [vector]: the number of subgroups. The default value is \code{2:4}.
\item \code{mindat} [scalar]: the minimum number of observations at each subgroup. The default value is 5\% of data.
\item \code{type} [character]: a type of optimal subgroup selection algorithm. At this stage, we offer two options. The option "perm" utilizes permutation test, while "NULL" passes a selection algorithm.
\item $\ldots$ [S4 class \code{kapsOptions}]: a list of minor parameters.
\end{itemize}

The primary arguments used for analysis are \code{formula} and \code{data}. 
All of the information created by \code{kaps()} is stored into an object from the \code{kaps} \code{S4} class. 
The output structure is given in Table \ref{slots}. 
In addition, five generic functions are available for the class: \code{show-method}, \code{print-method}, \code{plot-method}, \code{predict-method} and \code{summary-method}.

\begin{table}
\centering
\caption{The main slots for the \code{kaps} \code{S4} class.}
\label{slots}
\begin{tabular}{lll}
  \hline
  Slot & Type & Description \\
  \hline
  call & language & evaluated function call\\
  formula & Formula & formula to be used \\
  data & data.frame & data to be used in the model fitting \\
  groupID & vector & subgroup classified \\
  index & vector & an index for the selected K \\
  split.pt & vector & cut-off points selected\\
  results & list & results for each $K$\\
  Options & kapsOptions & minor parameters to be used \\
  X & scalar & test statistic with the worst pair of subgroups\\
  Z & scalar & overall test statistic\\
  pair & numeric & selected pair of subgroups \\
  \hline
\end{tabular}
\end{table}

\subsection*{Illustrative example}
To illustrate the function \code{kaps} with various options, we use a simple artificial data, \code{toy}, which consists of 150 artificial observations of the survival time (\code{time}), its censoring status (\code{status}) and the number of metastasis lymph nodes (LNs) (\code{meta}) as a covariate.
The data can be called up from the package \pkg{kaps}:
\begin{Schunk}
\begin{Sinput}
R> library("kaps")
R> data(toy)
R> head(toy)
\end{Sinput}
\begin{Soutput}
  meta status time
1    1      0    0
2    4      1   26
3    0      1   22
4    9      1   15
5    0      1   70
6    1      0   96
\end{Soutput}
\end{Schunk}

Here we utilize just 3 variables: \code{meta}, \code{status} and \code{time}. 
The number of metastasis LNs, \code{meta}, is used as an ordered prognostic factor for finding heterogeneous subgroups.
The available data have the following structure:

\begin{Schunk}
\begin{Sinput}
R> str(toy)
\end{Sinput}

\begin{Soutput}
'data.frame':	150 obs. of  3 variables:
 $ meta  : int  1 4 0 9 0 1 0 5 0 0 ...
 $ status: num  0 1 1 1 1 0 0 0 1 0 ...
 $ time  : num  0 26 22 15 70 96 97 10 32 127 ...
\end{Soutput}

\end{Schunk}

\subsubsection*{Selecting a set of cut-off points for given $K$}
Suppose we specify the number of subgroups in advance. 
For instance, $K=3$. 
To select an optimal set of two cut-off points when $K=3$, the function \code{kaps} is called via the following statements
\begin{Schunk}
\begin{Sinput}
R> fit1 <- kaps(Surv(time, status) ~ meta, data = toy, K = 3)
R> fit1 
\end{Sinput}
\begin{Soutput}
Call:
kaps(formula = Surv(time, status) ~ meta, data = toy, K = 3)

	K-Adaptive Partitioning for Survival Data

Samples= 150 				Optimal K=3 


Selecting a set of cut-off points:
      Xk df Pr(>|Xk|)  X1 df Pr(>|X1|) adj.Pr(|X1|) cut-off points  
K=3 36.8  2         0 7.2  1    0.0073     0.014701          0, 10 *
---
Signif. codes:  0 ‘***’ 0.001 ‘**’ 0.01 ‘*’ 0.05 ‘.’ 0.1 ‘ ’ 1 

P-values of pairwise comparisons
            0<=meta<=0 0<meta<=10
0<meta<=10       1e-04          -
10<meta<=38     <.0000     0.0073
\end{Soutput}
\end{Schunk}
On the \proglang{R} command, we first create an object \code{fit1} by the function \code{kaps()} with the three input arguments \code{formula}, \code{data}, and \code{K}. 
The object \code{fit1} has the \code{S4} class \code{kaps}. 
The function \code{show} returns the outputs of the object, consisting of three parts: \code{Call, Selecting a set of cut-off points}, and \code{P-values of pairwise comparisons}.

The first part, \code{Call}, displays the model formula with a dataset and a number for $K$. 
In this example, the prognostic factor, $meta$, is used to find three heterogeneous subgroups since $K=3$. 
Next, the information regarding the selection of an optimal set of cut-off points is provided for given $K$ in the table. 
In this part, the \code{Xk} ($T_{K-1}^{2}$) and \code{X1} ($T_{1}^{2}(s_{K}^*)$) mean the overall and minimum pairwise test statistics, and the \code{Pr(>|Xk|)} and \code{Pr(>|X1|)} denote their corresponding $p$-values.
The \code{adj.Pr(|X1|)} indicates a permuted $p$-value for the worst-pair with the smallest test statistic.

When $K=3$, an optimal set of two cut-off points selected by the algorithm is $s_K^{*} = \{0, 10 \}$. 
The two cut-off points are used to partition the data into three groups: $meta = 0$, $0 < meta \leq 10$, and $10 < meta \leq 38$. 
For the three subgroups, the overall test statistic $T_{K-1}^{2}$ (\code{Xk}), the degree of freedom (\code{df}), and the $p$-value (\code{Pr(|Xk|)}) are given.
Note that if $K$ is not significant, the output part is changed from \code{"Optimal K=3"} to \code{"Optimal K<3"}.
It means the value of the argument $K$ may be less than the present input value. 
Lastly,  the $p$-values of pairwise comparisons among all the pairs of subgroups are provided. 

The $p$-values can be adjusted for multiple comparison, as shown below.
\begin{Schunk}
\begin{Sinput}
R> fit2 <- kaps(Surv(time, status) ~ meta, data = toy, K=3,
+ p.adjust.methods = "holm")
R> fit2
\end{Sinput}

\begin{Soutput}
Call:
kaps(formula = Surv(time, status) ~ meta, data = toy, K = 3, 
    p.adjust.methods = "holm")

	K-Adaptive Partitioning for Survival Data

Samples= 150 				Optimal K=3 


Selecting a set of cut-off points:
      Xk df Pr(>|Xk|)  X1 df Pr(>|X1|) adj.Pr(|X1|) cut-off points  
K=3 36.8  2         0 7.2  1    0.0073     0.012101          0, 10 *
---
Signif. codes:  0 ‘***’ 0.001 ‘**’ 0.01 ‘*’ 0.05 ‘.’ 0.1 ‘ ’ 1 

P-values of pairwise comparisons
            0<=meta<=0 0<meta<=10
0<meta<=10       2e-04          -
10<meta<=38     <.0000     0.0073
\end{Soutput}
\end{Schunk}
It is based on the internal function \code{p.adjust}. 
The default value of \code{p.adjust.methods} is "none". 
The only difference between the objects \code{fit1} and \code{fit2} is the $p$-values of pairwise comparisons. 
For more information, refer to the help page of the function \code{p.adjust}. 
The Kaplan-Meier survival curves can be obtained by

\begin{Schunk}
\begin{Sinput}
R> plot(fit1)
\end{Sinput}
\end{Schunk}

\begin{figure}
   \centering
  \includegraphics[width= 8cm]{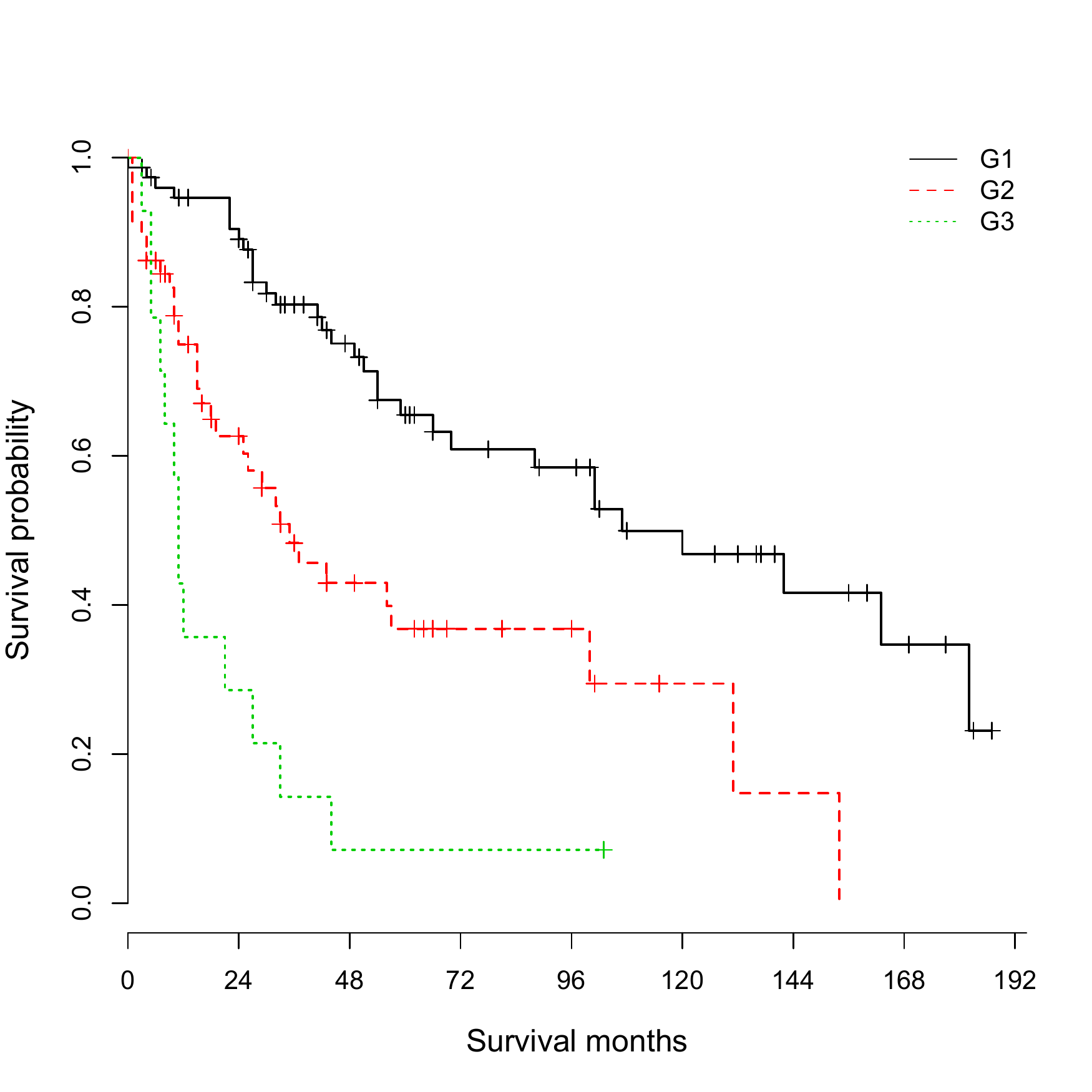}
  \caption{Kaplan-Meier survival curves for the \code{toy} dataset with three subgroups: \code{G1}= \{$meta = 0$\}, \code{G2}= \{$0 < meta \leqq 10$\}, and \code{G3}= \{$10 < meta \leqq 38$\}.}
  \label{toy:km}
\end{figure}

It provides Kaplan-Meier survival curves for the selected subgroups as seen in Figure \ref{toy:km}. 
The method \code{summary} shows the tabloid information for the subgroups. 
It consists of the number of observations (\code{N}), the survival median time (\code{Med}), and the 1-year (\code{yrs.1}), 3-year (\code{yrs.3}), and 5-year (\code{yrs.5}) survival times. 
The rows mean orderly for all the data (\code{All}) and each subgroup.

\begin{Schunk}
\begin{Sinput}
R> summary(fit1) 
\end{Sinput}

\begin{Soutput}
          N Med yrs.1 yrs.3 yrs.5
All     150  57 0.813 0.609 0.488
Group=1  76 107 0.946 0.803 0.655
Group=2  60  35 0.749 0.456 0.368
Group=3  14  11 0.357 0.143 0.000
\end{Soutput}
\end{Schunk}

\subsubsection*{Finding an optimal $K$}

The number ($K$) of subgroups is usually unknown and may not therefore be specified in advance. 
Rather, an optimal $K$ can be selected by the algorithm for a given range of $K$ as follows:

\begin{Schunk}
\begin{Sinput}
R> fit3 <- kaps(Surv(time, status) ~ meta, data = toy, K = 2:4)
R> fit3
\end{Sinput}

\begin{Soutput}
Call:
kaps(formula = Surv(time, status) ~ meta, data = toy, K = 2:4)

	K-Adaptive Partitioning for Survival Data

Samples= 150 				Optimal K=3 


Selecting a set of cut-off points:
      Xk df Pr(>|Xk|)    X1 df Pr(>|X1|) adj.Pr(|X1|) cut-off points    
K=2 26.4  1         0 26.37  1    0.0000      0.00000              8 ***
K=3 36.8  2         0  7.20  1    0.0073      0.01240          0, 10 *  
K=4 38.0  3         0  1.89  1    0.1692      0.16752        0, 3, 6    
---
Signif. codes:  0 ‘***’ 0.001 ‘**’ 0.01 ‘*’ 0.05 ‘.’ 0.1 ‘ ’ 1 

P-values of pairwise comparisons
            0<=meta<=0 0<meta<=10
0<meta<=10       1e-04          -
10<meta<=38     <.0000     0.0073
\end{Soutput}
\end{Schunk}

Optimal sets of cut-off points are selected for each $K$, as seen in the output with the title "\code{Selecting a set of cut-off points}". 
The explanation for the output is the same as that of the previous subsection. 
Then an optimal $K$ is selected by the algorithm with permutation test as described in Section 2.2, respectively. 
In the output, \code{Xk} and \code{X1} indicate the overall and worst-pair test statistics.
Their degrees of freedom and $p$-values are followed in the output. 
The \code{"adj.Pr(|X1|)"} is the Bonferroni corrected permuted $p$ value for the worst pair by which we make a decision for the optimal $K$.
In this example,  an optimal $K$ is 3 because the worst pairs of comparisons were significant with significance level $\alpha = 0.05$ when $K = 2 \mbox{ and } 3$, and the worst-pair $p$-value for $K=4$ is rapidly increased.

\begin{figure}
   \centering
  \includegraphics[width= 14cm]{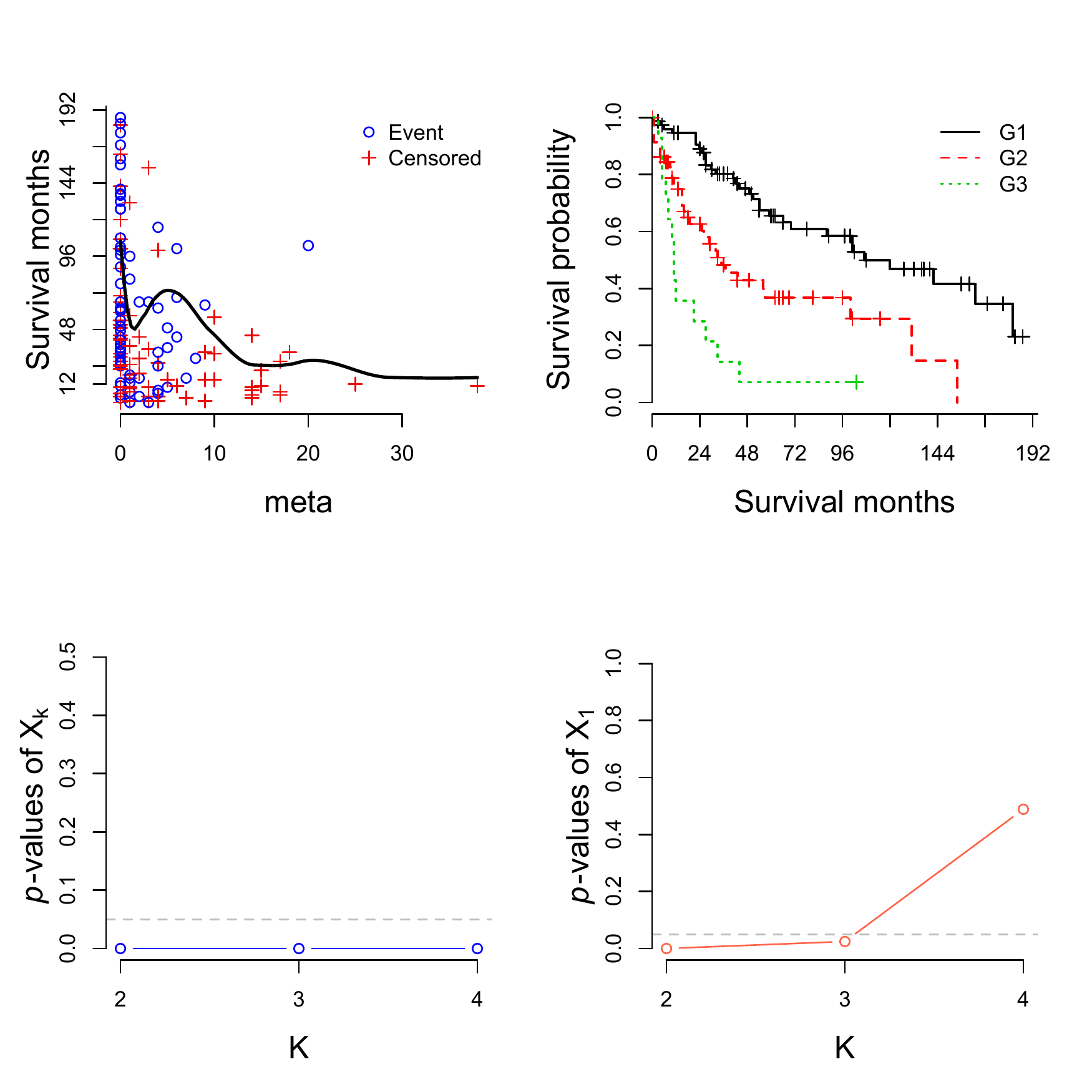}
  \caption{The top left panel is the scatter plot of survival times against the prognostic factor with the line fitted by local censored regression. The top right panel is the Kaplan-Meier survival curves for the selected subgroups. The panels at the bottom are the plots of the overall and worst-pair $p$-values against $K$ with significance level $\alpha = 0.05$. }
  \label{toy:plot}
\end{figure}

The test statistic for determining an optimal $K$ can be displayed by
\begin{Schunk}
\begin{Sinput}
R> plot(fit3)
\end{Sinput}
\end{Schunk}
It generates the four plots shown in Figure \ref{toy:plot}. 
The top left panel is the scatterplot of survival times against the prognostic factor $meta$ with the line fitted by local censored regression \citep{LRL}. 
The top right panel is the Kaplan-Meier survival curves for the subgroups selected with the optimal $K$. 
At the bottom are displayed the plots of the overall and worst-pair $p$ values against $K$. 
The dotted lines indicate thresholds for significance ($\alpha = 0.05$).

The outputs for $K$s can also be printed out. 
For instance, when $K$ is 4, the output is printed out as follows.
\begin{Schunk}
\begin{Sinput}
R> print(fit3, K= 4)
\end{Sinput}

\begin{Soutput}

P-values of pairwise comparisons when K = 4  

           0<=meta<=0 0<meta<=3 3<meta<=6
0<meta<=3       1e-04         -         -
3<meta<=6      0.2812    0.1687         -
6<meta<=38     <.0000    0.1151    0.0092
\end{Soutput}
\end{Schunk}
It gives information about pairwise comparisons for a specific $K$.

\subsection*{System requirements, availability and installation}
\pkg{kaps} is an \proglang{R} package developed by employing the following \proglang{R} packages: \pkg{methods}, \pkg{survival}, \pkg{Formula} and \pkg{coin}. 
It requires \proglang{R} ($>$3.0.0) and runs under Windows and Unix like operating systems. 
The source code of development version and detailed installation guide for \pkg{kaps} are freely available under the terms of GNU license from \url{https://sites.google.com/site/sooheangstat/}. 
The stable version of \pkg{kaps} is also available at the Comprehensive R Archive Network (\url{http://cran.r-project.org/package=kaps}).\\

\begin{tabular}{ll}
Project name         & $K$-Adaptive Partitioning for Survival Data\\
Operating system(s)  & Platform independent \\
Other requirements   & None\\
Programming language & R ($\geq$3.0.0)\\
License              & GNU GPL version 3\\
\end{tabular}\\

\end{document}